\documentclass{aa}
\usepackage{graphicx,natbib,amsmath,amssymb,multirow}
\usepackage[T1]{fontenc}
\usepackage{mathptmx}
\usepackage[colorlinks=true,citecolor=blue,urlcolor=blue]{hyperref}
\bibpunct{(}{)}{;}{a}{}{,}
\newcommand{\kms}{\ifmmode{~{\rm km\,s}^{-1}}\else{~km~s$^{-1}$~}\fi}
\newcommand{\msun}{\ifmmode{{\rm M}_{\odot}}\else{${\rm M}_{\odot}$~}\fi}
\newcommand{\msunit}{\ifmmode{M_{\odot}}\else{$M_{\odot}$~}\fi}
\newcommand{\lsun}{\ifmmode{{\rm L}_{\odot}}\else{${\rm L}_{\odot}$}\fi}
\defcitealias{vaytet2012}{Paper~I}
\defcitealias{masunaga2000}{MI00}
\defcitealias{tomida2013}{Tea13}
\defcitealias{saumon1995}{SCvH95}
\defcitealias{stamatellos2007}{SWBG07}

\begin{document}

\title{Simulations of protostellar collapse using multigroup radiation hydrodynamics. II. The second collapse}
\titlerunning{Protostellar collapse using multigroup RHD. II.}

\author{Neil Vaytet$^{1}$, Gilles Chabrier$^{1,2}$, Edouard Audit$^{3,4}$, Beno\^{i}t Commer\c{c}on$^{5}$, Jacques Masson$^{1}$, Jason Ferguson$^{6}$ and Franck Delahaye$^{7}$}
\authorrunning{Vaytet et al.}

\institute{$^{1}$ \'{E}cole Normale Sup\'{e}rieure de Lyon, CRAL, UMR CNRS 5574, Universit\'{e} Lyon I, 46 All\'{e}e d'Italie, 69364 Lyon Cedex 07, France\\
           $^{2}$ School of Physics, University of Exeter, Exeter, EX4 4QL, UK\\
           $^{3}$ Maison de la Simulation, USR 3441,  CEA - CNRS - INRIA - Universit\'{e} Paris-Sud - Universit\'{e} de Versailles, 91191 Gif-sur-Yvette, France\\
           $^{4}$ CEA/DSM/IRFU, Service d'Astrophysique, Laboratoire AIM, CNRS, Universit\'{e} Paris Diderot, 91191 Gif-sur-Yvette, France\\
           $^{5}$ Laboratoire de radioastronomie, (UMR CNRS 8112), \'{E}cole normale sup\'{e}rieure et Observatoire de Paris, 24 rue Lhomond, 75231 Paris Cedex 05, France\\
           $^{6}$ Department of Physics, Wichita State University, Wichita, KS 67260-0032, USA\\
           $^{7}$ LERMA, Observatoire de Paris, ENS, UPMC, UCP, CNRS, 5 Place Jules Janssen, 92190 Meudon, France
          }

\date{Received / Accepted}

\offprints{neil.vaytet@ens-lyon.fr} 

\abstract{Star formation begins with the gravitational collapse of a dense core inside a molecular cloud. As the collapse 
progresses, the centre of the core begins to heat up as it becomes optically thick. The temperature and density in the centre 
eventually reach high enough values where fusion reactions can ignite; the protostar is born. This sequence of events entail many 
physical processes, of which radiative transfer is of paramount importance. Simulated collapsing cores without radiative transfer 
rapidly become thermally supported before reaching high enough temperatures and densities, preventing the formation of 
stars.}
{Many simulations of protostellar collapse make use of a grey treatment of radiative transfer coupled to the hydrodynamics. However,
interstellar gas and dust opacities present large variations as a function of frequency, which can potentially be overlooked by
grey models and lead to significantly different results. In this paper, we follow-up on a previous paper on the collapse and 
formation of Larson's first core using multigroup radiation hydrodynamics (Paper~I) by extending the calculations to the second 
phase of the collapse and the formation of Larson's second core.}
{We have made the use of a non-ideal gas equation of state as well as an extensive set of spectral opacities in a spherically
symmetric fully implicit Godunov code to model all the phases of the collapse of a 0.1, 1 and 10~\msun cloud cores.}
{We find that, for a same central density, there are only small differences between the grey and multigroup simulations. The first
core accretion shock remains supercritical while the shock at the second core border is found to be strongly subcritical with all
the accreted energy being transfered to the core. The size of the first core was found to vary somewhat in the different simulations
(more unstable clouds form smaller first cores) while the size, mass and temperature of the second cores are independent of initial
cloud mass, size and temperature.}
{Our simulations support the idea of a standard (universal) initial second core size of $\sim 3 \times 10^{-3}$ AU and mass
$\sim 1.4 \times 10^{-3}$ \msun. The grey approximation for radiative transfer appears to perform well in one-dimensional simulations
of protostellar collapse, most probably because of the high optical thickness of the majority of the protostar-envelope system. A
simple estimate of the characteristic timescale of the second core suggests that the effects of using multigroup radiative transfer
may be more important in the long term evolution of the proto-star.}


\maketitle

\section{Introduction}

The formation of new low-mass stars begins with the gravitational collapse of a cold dense core inside a molecular cloud which then
heats up in its centre  as the pressure and density increase from the compression, a problem which entails many physical processes
(hydrodynamics, radiative  transfer, magnetic fields, etc...) over a very large range of spatial scales
\citep{larson1969,stahler1980,masunaga1998}. The collapsing material is initially optically thin to the thermal emission from the cold 
gas and dust grains and all the energy gained from compressional heating is transported away by the escaping radiation, which causes 
the cloud to collapse isothermally in the initial stages of the formation of a protostar. When the optical depth of the cloud reaches 
unity, the radiation is absorbed by the system which starts heating up, taking the core collapse through its adiabatic phase. The 
strong compression forms an accretion shock at the border of the adiabatic core (also known as Larson's first core). The first core 
continues to accrete the surrounding material, grows in mass but still contracts further due to the gravity which overcomes the 
thermal supportas well as radiative losses. With contraction comes a rise in gas temperature, and as it reaches 2000 K, the hydrogen molecules begin to 
dissociate. This leads the system into its second phase of collapse because of the endothermic nature of the dissociation process. The 
second collapse ends when most or all of the $\text{H}_{2}$ molecules have been split and a second much more dense and compact 
hydrostatic core is formed at the centre; Larson's second core is born (\citealt{larson1969}; \citealt{masunaga2000}, hereafter
\citetalias{masunaga2000}; \citealt{stamatellos2007}, hereafter \citetalias{stamatellos2007}; \citealt{tomida2013}, hereafter
\citetalias{tomida2013}).

Numerical studies of the first and second collapse are very demanding, they require the solutions to the full radiation hydrodynamics 
(RHD) system of equations, and three-dimensional RHD simulations have only just recently become possible with modern computers. One-dimensional
studies including the most complex physics are still leading in terms of understanding and discovering the physical 
processes at work. In particular, including frequency dependent radiative transfer is essential
to properly take into account the strong variations of the interstellar gas and dust opacities as a function of frequency 
\citep[see for example][]{ossenkopf1994,li2001,draine2003a,semenov2003,ferguson2005}. Three-dimensional full frequency-dependent radiative transfer
is still out of reach of current computer architectures and only rare attempts with simplified methods have been made in the context of star
formation \citep[see][for instance]{kuiper2011}. In this paper, we continue the recent 1D simulations of the 
first collapse of \citet{vaytet2012} \citepalias[hereafter][]{vaytet2012} by following the evolution of the system through the second 
phase of the collapse up to the formation of the second Larson core. This involves the inclusion of a sophisticated equation of state (EOS) to
reproduce the effects of the $\text{H}_{2}$ dissociation, which cannot be achieved using the more common ideal gas EOS. A new set of 
frequency-dependent opacities was also developed since the one used in \citetalias{vaytet2012} was only valid for temperatures below 
$\sim 2000$ K.

We first describe the numerical method, EOS and opacities used in the simulations. The frequency dependence is implemented through the 
multigroup method in which the frequency domain is divided into a finite number of bins or groups, and the opacities are averaged 
within each group \citep{vaytet2011}. The thermal evolution of the system is then described and radial profiles are presented. 
Simulations of the collapse of clouds with different initial masses were performed and the properties of the first and second cores 
are listed.

\section{The multigroup RHD collapse simulations}\label{sec:simulations}

\subsection{Numerical method and initial conditions}\label{sec:init_cond}

The code used to solve the multigroup RHD equations was an updated version of the 1D fully implicit Lagrangean code used in 
\citetalias{vaytet2012}. The vast majority of the code was unchanged but a new EOS and opacity database were added (see
section~\ref{sec:eos}), as well as a parallelisation scheme using OpenMP. The grid comprises 2000 cells logarithmically spaced in the
radial direction.

The initial setup for the dense core collapse was identical to \citetalias{vaytet2012}. A uniform density sphere of mass
$M_{0} = 1~\msun$, temperature $T_{0} = 10$ K ($c_{s0} = 0.187 \kms$) and radius $R_{0} = 10^{4}$ AU collapses under its own gravity. 
The ratio of thermal to gravitational energy in the cold gas cloud is
\begin{equation}\label{equ:energy_ratios}
\alpha = \frac{5R_{0}k_{B}T_{0}}{2GM_{0}\mu m_{H}} = 1.02
\end{equation}
where $G$ is the gravitational constant, $k_{B}$ is the Boltzmann constant, $\mu$ is the mean molecular weight and $m_{H}$ is the mass
of the hydrogen atom. The cloud's free-fall time is $t_{\text{ff}} \sim 0.177$ Myr.\footnote{Note that in \citetalias{vaytet2012}, we
had $\alpha = 0.98$ for the same set of initial conditions because of a mean particle weight of 2.375 which corresponds to a gas of
solar abundances, while in our new EOS, only H and He are considered and the mean particle weight is 2.31 for a He concentration of
0.27.} The radiation temperature is in equilibrium with the gas temperature (the  energy of a black body with $T = 10$ K is divided
among the frequency groups according to the Planck distribution) and the radiative  flux is set to zero everywhere. The boundary
conditions are reflexive at the centre of the grid ($r = 0$) and have imposed values equal to the initial conditions at the outer edge
of the sphere.

\subsection{Gas equation of state}\label{sec:eos}

We used the gas equation of state (EOS) of \citet[hereafter \citetalias{saumon1995}]{saumon1995} which models the thermal properties of a gas containing the following
species: $\text{H}_{2}$, H, $\text{H}^{+}$, He, $\text{He}^{+}$ and $\text{He}^{2+}$ (the He mass concentration was 0.27). We extended
the original EOS table to low temperatures (below 125 K) and densities (below $10^{-6}~\text{g~cm}^{-3}$) by computing the partition
function for H, He and $\text{H}_{2}$ (taking into account the correct rotational excitation levels for $\text{H}_{2}$). The
Debye-H\"{u}ckel interaction term and the \citet{hummer1988} excluded volume interaction are both included in the computation of the
chemical equilibrium, and the zero point of energy is chosen as the ground state of the $\text{H}_{2}$ molecule. The calculation is
trivial but rather tedious and, for the sake of conciseness, is not explicited further.

\begin{figure}[!ht]
\centering
\includegraphics[scale=0.58]{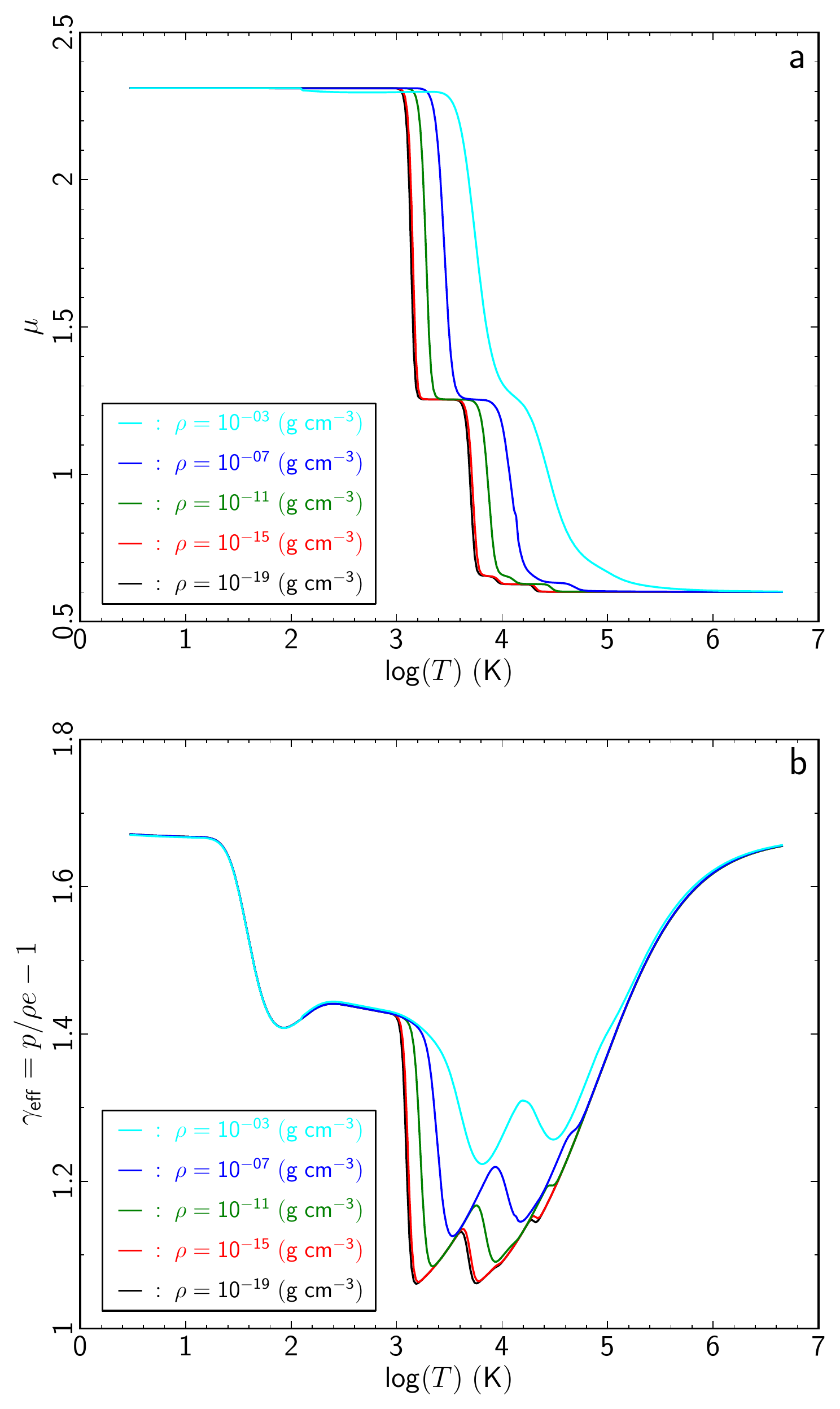}
\caption[EOS table]{The \citetalias{saumon1995} EOS and its extension to low densities: $\mu$ (a) and $\gamma_{\text{eff}}$ (b) as a
function of temperature for five different densities (see colour key in each panel).}
\label{fig:eos}
\end{figure}

Figure~\ref{fig:eos} displays $\mu$ and the effective ratio of specific heats ($\gamma_{\text{eff}}$) obtained from the resulting table
as a function of temperature for five different gas densities. The table recovers the transition around 85 K
\citep[see][p.~378]{sears1975} from a monatomic $\gamma_{\text{eff}} = 5/3$ at low temperatures (when the $\text{H}_{2}$
rotational levels are frozen) to a diatomic gas with $\gamma_{\text{eff}} = 7/5$. Due to the Boltzmannian nature of the energy
distribution, rotational levels start to get excited as early as 30 K, and the transition from a monatomic to a diatomic
$\gamma_{\text{eff}}$ operates smoothly as the temperature increases. This realistic EOS table also enables us to properly model the
second phase of the collapse which begins with the dissociation of $\text{H}_{2}$ around $T\sim2000$ K (also visible in
Fig.~\ref{fig:eos}; note that the dissociation temperature varies somewhat with the gas density).

\subsection{Interstellar dust and gas opacities}\label{sec:opacities}

\begin{figure*}[!ht]
\centering
\includegraphics[scale=0.5]{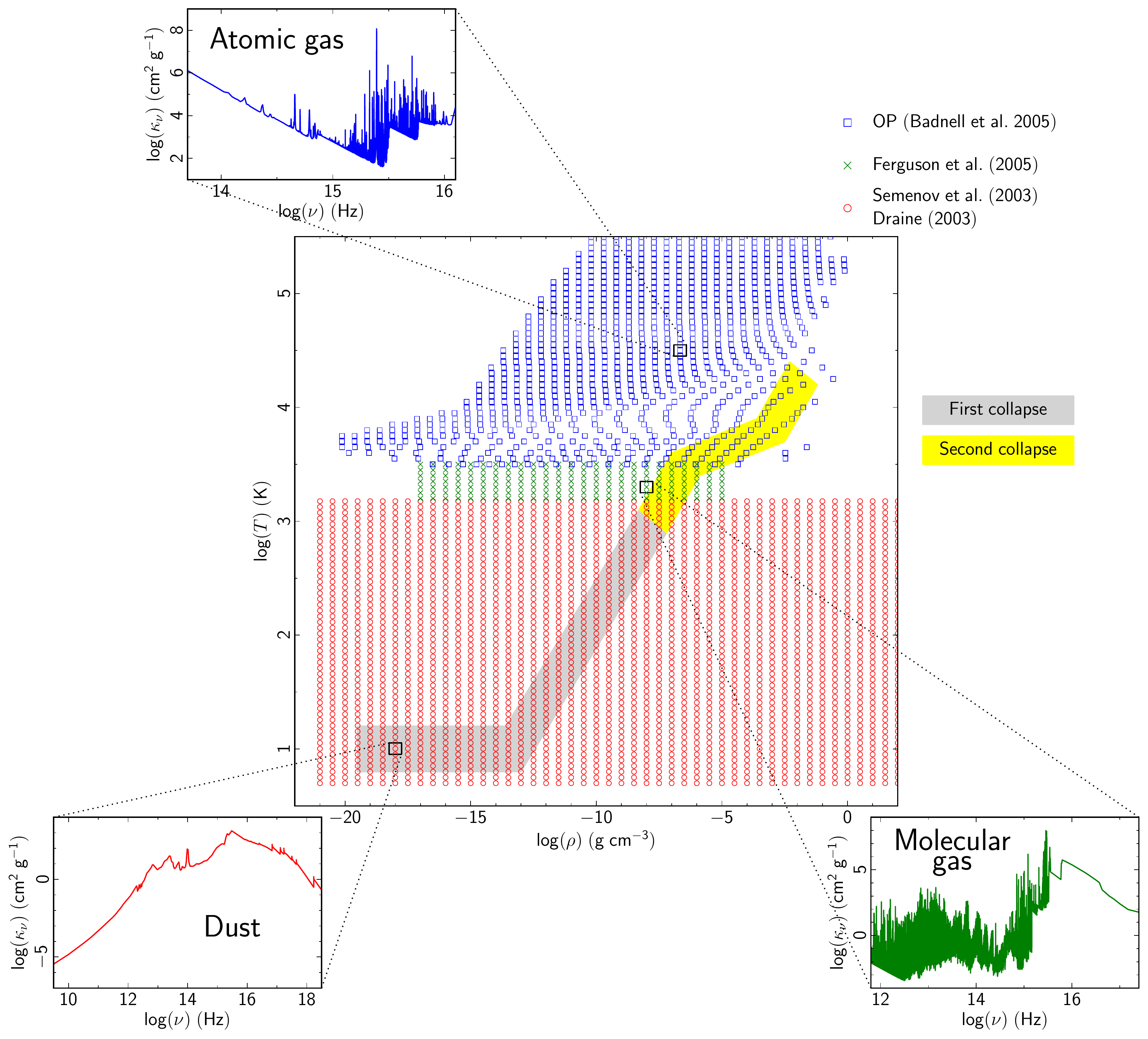}
\caption[Spectral opacities in the $(\rho,T)$ plane]{Spectral opacities $\kappa(\rho,T,\nu)$. Each data point represents a set of 
spectral opacities for a given density and temperature . We have compiled the table from three different opacity collections. At
low temperatures (below 1500 K), we have taken the dust opacities from \citet{semenov2003} which we have completed by the dust
grains opacities of \citet{draine2003b} at the high frequency end (red circles). For temperatures between 1500 and 3200 K, the
opacities for molecular gas based on the \citet{ferguson2005} calculations were used (green crosses). Finally, for temperatures
above 3200 K, we adopted the OP atomic gas opacities \citep[blue squares]{badnell2005}. Two coloured areas, grey and yellow, show
the approximate range of densities and temperatures typically reached during the first and second stages of the collapse of a
cloud core, respectively.}
\label{fig:kappaRTmap}
\end{figure*}

For our multigroup simulations, we require temperature and density-dependent monochromatic opacities for the interstellar gas. A 
complete set of monochromatic opacities, covering the range $10^{-19}~\text{g~cm}^{-3} < \rho < 10^{2}~\text{g~cm}^{-3}$ and 
$5~\text{K} < T < 10^{7}~\text{K}$ does not exist in the literature and we have had to piece together several different tables. We 
used three different opacity sets, one for interstellar dust, one for molecular gas and one for atomic gas; this is illustrated in 
Fig.~\ref{fig:kappaRTmap}.

\begin{figure}[!ht]
\centering
\includegraphics[scale=0.91]{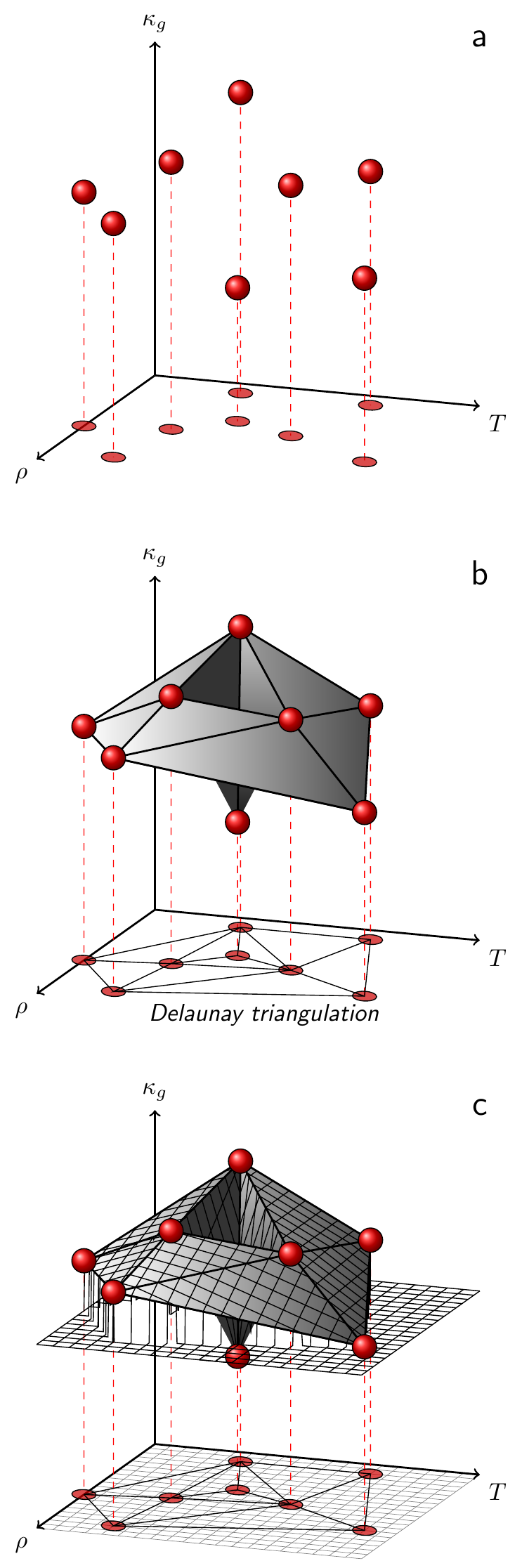}
\caption[Construction of the opacity table]{The three steps in the construction of the regular opacity table. (a) Group average 
opacities are computed for each spectral point in the $(\rho,T)$ table. (b) A Delaunay triangulation is computed in the $(\rho,T)$ 
plane using each table point as a triangle vertex. Each triangle from the Delaunay triangulation represents a plane in the
$(\rho,T,\kappa_{g})$ space. (c) We then overlay a fine rectanglar grid of opacity points which are computed from their coordinates in 
the opacity planes (see text for more details).}
\label{fig:kappagrid}
\end{figure}

At low temperatures (below 1500 K), the opacities of the interstellar material are dominated by the one percent in mass of dust grains
present in the medium. For this temperature region, we used \citepalias[as in][]{vaytet2012} the monochromatic opacities from 
\citet{semenov2003}\footnote{\url{http://www.mpia.de/homes/henning/Dust_opacities/Opacities/opacities.html}} who provide dust 
opacities for various types of grains in five different temperature ranges (we assume that the dust opacities are independent of gas 
density and that the dust is in thermal equilibrium with the gas; see \citealt{galli2002} for example). We used the spectral opacities 
for homogeneous spherical dust grains and normal iron content in the silicates ($\text{Fe}/(\text{Fe} + \text{Mg})=0.3$). At the high 
frequency end ($\nu > 3\times10^{15}$ Hz), we have completed the set with high-energy dust opacities from \citet{draine2003b}, giving 
a total of 583 frequency bins between $3\times10^{9}$ and $3\times10^{18}$ Hz. The opacities at those high frequencies are not very 
important since the dust is only present at low gas temperatures (below 1500 K) and there will be virtually no radiative energy in 
that part of the spectrum. The Draine opacities were only included so that the UV and X-ray frequency groups had a non-zero opacity in 
the cold parts of the simulation. The dust $\kappa(\rho,T,\nu)$ data set is pictured in Fig.~\ref{fig:kappaRTmap} (red circles).

For temperatures between $\sim 1500 - 3200$ K, the dust grains are rapidly destroyed and molecular gas opacities prevail \citep[see][]
{ferguson2005}. We have used a set of monochromatic opacities for the range
$10^{-17}~\text{g~cm}^{-3} < \rho < 10^{-5}~\text{g~cm}^{-3}$ and $1500~\text{K} < T < 3200~\text{K}$ for an interstellar gas with 
solar abundances comprising $\sim$ 26,000 frequency bins between $6\times10^{11}$ and $3\times10^{17}$ Hz, which were calculated based
upon the computations discussed in \citet{ferguson2005}. This is shown in Fig.~\ref{fig:kappaRTmap} (green crosses).

Finally, at temperatures above $\sim3200$ K, there are no more molecules and the atomic opacities take over. In the range 
$10^{-20}~\text{g~cm}^{-3} < \rho < 10^{6}~\text{g~cm}^{-3}$ and $10^{3.5}~\text{K} < T < 10^{8}~\text{K}$, we have used a set of 
monochromatic OP opacities with 10,000 frequency bins in the range $0.1 < h\nu/k_{B}T < 20$ \citep{badnell2005}. The atomic gas
opacities are represented in Fig.~\ref{fig:kappaRTmap} by blue squares.

Our resulting raw table covers the entire evolutionary track of a two-stage cloud collapse, as opposed to the one used by
\citetalias{tomida2013} which is incomplete, especially towards the high temperatures and densities reached during the second
collapse, where they used simple extrapolations as opposed to real opacities (see their appendix B).

\begin{figure}[!ht]
\centering
\includegraphics[scale=0.4]{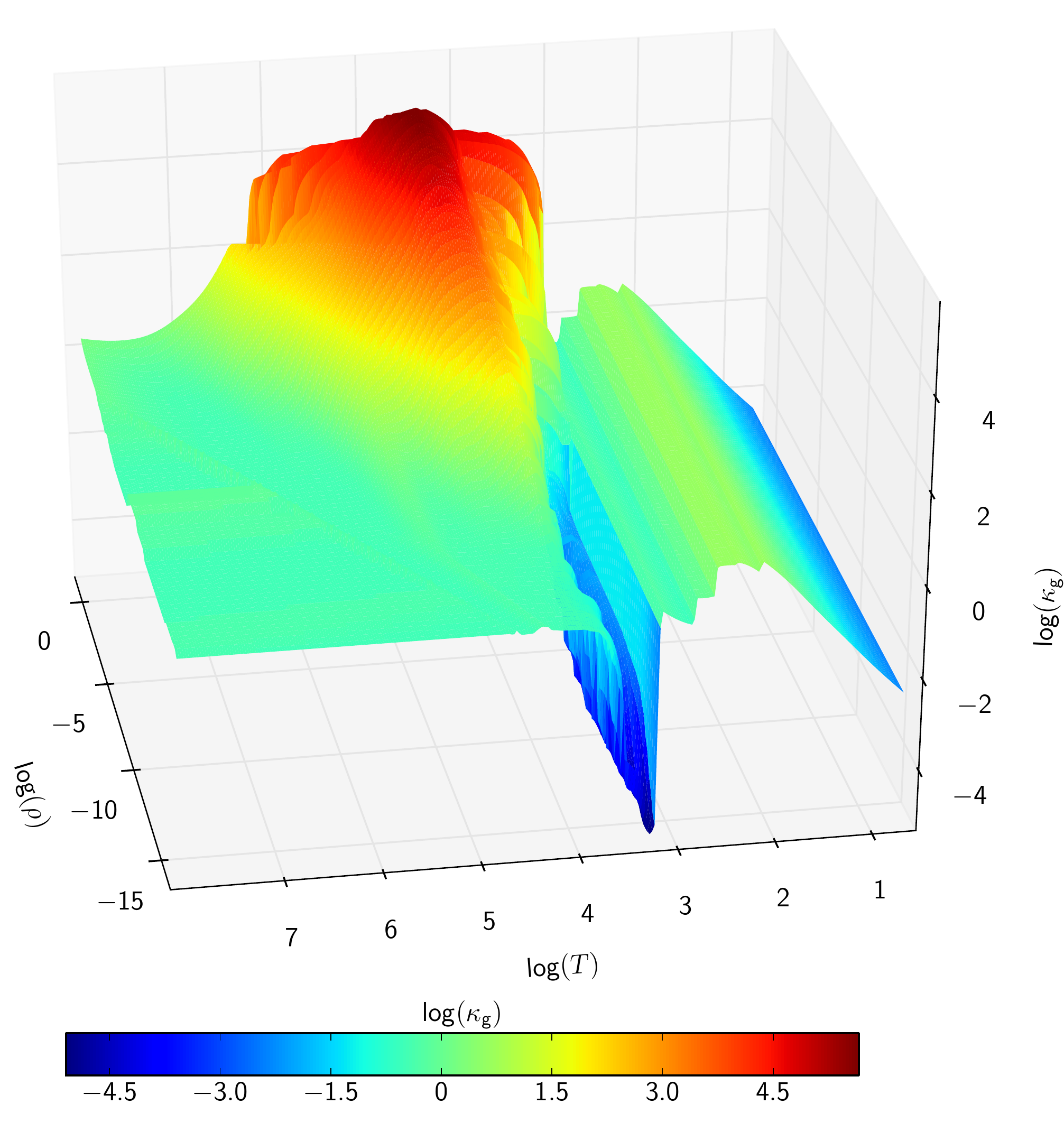}
\caption[Rosseland mean opacity]{Grey Rosseland mean opacity as a function of temperature and density. The steep trench (dark blue) 
along the density dimension around $T\sim 1500$ K corresponds to the destruction of dust grains, while the high mount (in red) 
represents the region where the dominant atomic gas opacities become very high.}
\label{fig:kappagrey}
\end{figure}

In our multigroup method, we need to compute Planck ($\kappa_{P}$) and Rosseland ($\kappa_{R}$) mean opacities as a function of density
and temperature for each  frequency group. We describe below the three-step process we employ to efficiently compute the group mean
opacities $\kappa_{Pg}$ and $\kappa_{Rg}$ (illustrations in Fig.~\ref{fig:kappagrid}).
\begin{itemize}
\item[(a)] $\kappa_{Pg}$ and $\kappa_{Rg}$ are computed for each spectral point in the $(\rho,T)$ table (Fig.~\ref{fig:kappaRTmap}) 
once at the beginning of the simulation.
\item[(b)] Since the points in the table are not all regularly spaced in $\rho$ and $T$, a Delaunay triangulation is computed in the 
$(\rho,T)$ plane using each table point as a triangle vertex.
\item[(c)] Each triangle from the Delaunay triangulation represents a plane in the $(\rho,T,\kappa_{g})$ space. We are now able to 
overlay a fine rectanglar grid of opacity points which are computed from their coordinates in the opacity planes (triangles).
\end{itemize}
This fine rectangular mean opacity grid allows for fast index finding and efficient bicubic interpolation during the rest of the 
simulation. The edges of the table, outside of all the triangles, were filled by simply using the outermost value of the closest
triangle, giving more or less a flat opacity surface. This was only included for consistency as these extreme values of $(\rho,T)$
were never reached in the simulations. The fine regular mesh of Rosseland mean opacities averaged over the entire frequency range
(in the case of a single frequency group) is shown in Fig.~\ref{fig:kappagrey}. We note the presence of the sharp opacity gap in
the region $\log(T)\sim3.2$ corresponding to the destruction of the dust grains. We also see that there is an opacity peak for
$\log(T)\sim 4-5$ at high densities.

\section{Results}\label{sec:results}

\begin{figure}[!ht]
\centering
\includegraphics[scale=0.40]{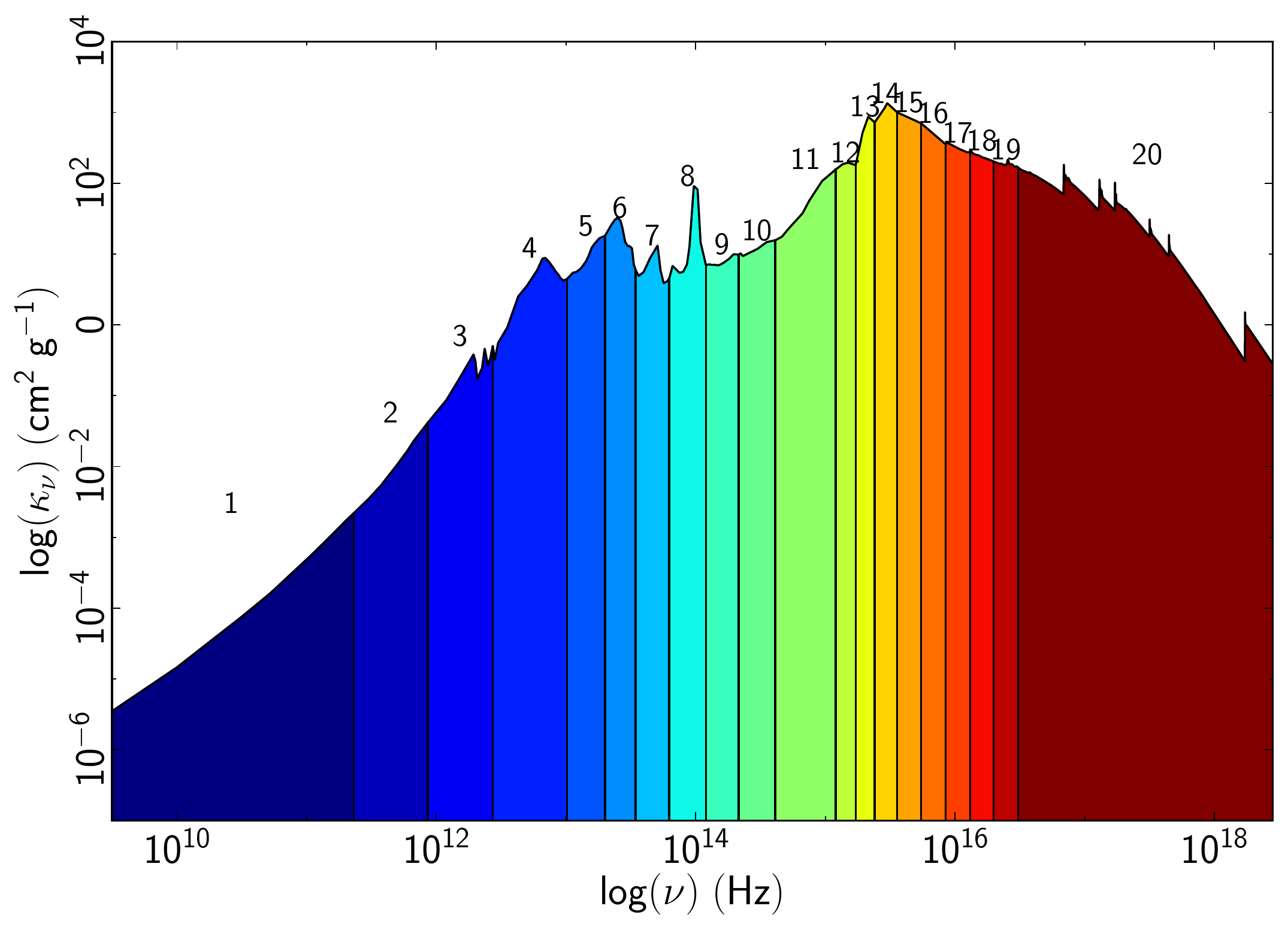}
\caption[Frequency domain decomposition]{Decomposition of the frequency domain using 20 groups, presented over the monochromatic dust 
opacities for illustration purposes. The first and last groups are used to make sure no energy is omitted at the low and high 
ends of the spectrum, respectively. The other groups offer an almost log-regular splitting of frequencies in the range 
$2.0\times10^{11} - 3\times10^{16}$ Hz. The group numbers are indicated just above the opacity curve. The presented spectral opacities
are for typical initial conditions of $\rho=10^{-18}~\text{g~cm}^{-3}$ and $T = 10$ K.}
\label{fig:kappanu}
\end{figure}

A grey (run 1) and a multigroup (run 2) simulations of the collapse of a dense core were performed (see Table~\ref{tab:simulation-params}).
In the grey run, the radiative quantities were integrated over the entire frequency range (0 to $10^{19}$ Hz) while for the multigroup run,
20 frequency groups were used; the decomposition of the frequency domain is illustrated in Fig.~\ref{fig:kappanu}. The simulations were run
until the central density reached $\rho_{c} = 6 \times 10^{-2}~\text{g~cm}^{-3}$.

\subsection{The two phases of the 1 \msunit cloud collapse}\label{sec:secondcollapsephases}

\begin{figure*}[!ht]
\centering
\includegraphics[scale=0.35]{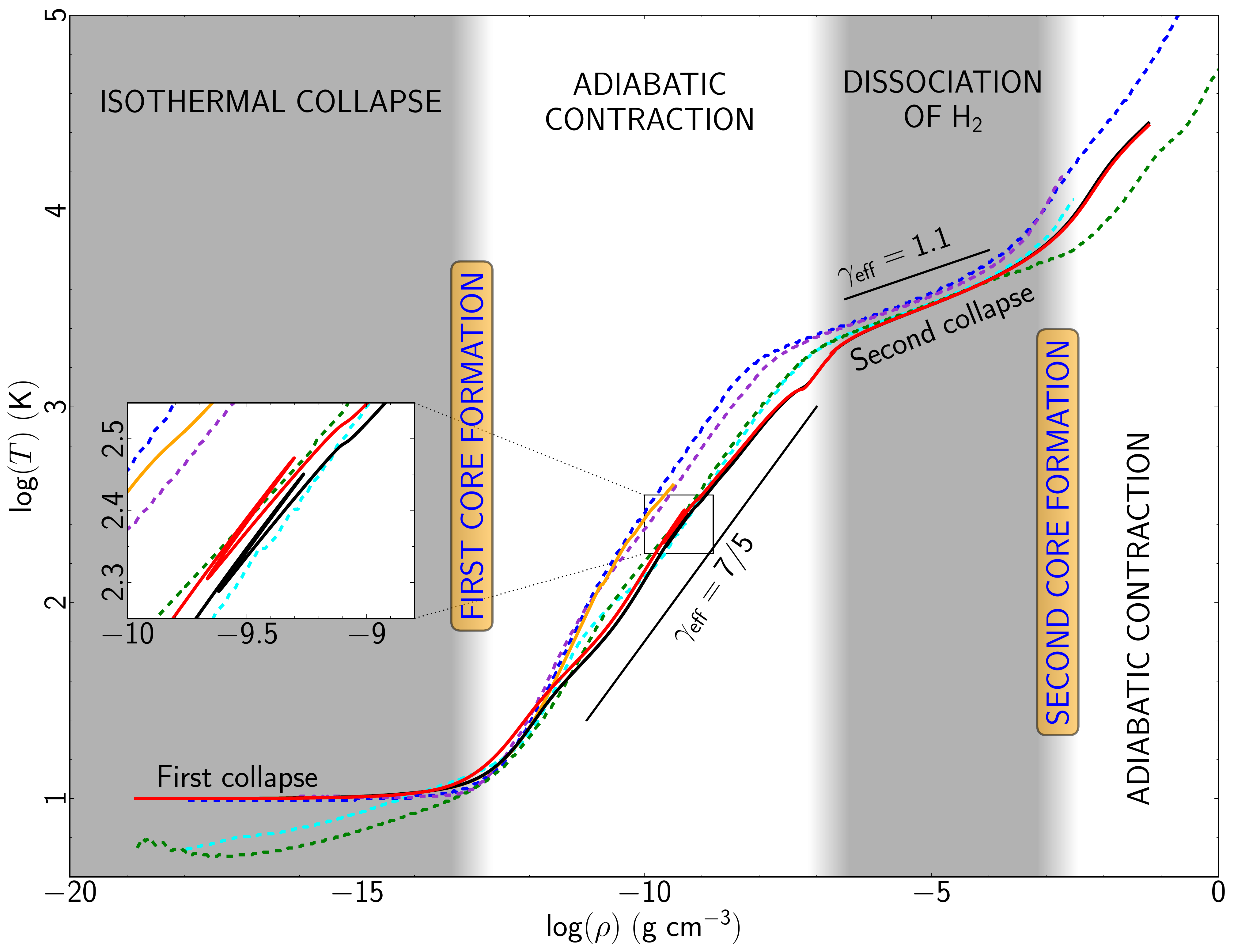}
\caption[Centre thermal evolution]{Thermal evolution at the centre of the collapsing cloud: the central temperature as a function of 
the central density for the grey (solid black) and multigroup (solid red) simulations. The different phases of the protostellar 
collapse are labeled. The results from \citetalias{masunaga2000} (dashed green), \citet{whitehouse2006} (dashed purple),
\citetalias{stamatellos2007} (dashed cyan) and \citetalias{tomida2013} (dashed blue) are also plotted for comparison. The orange
solid line is a grey simulation using an ideal gas EOS.}
\label{fig:rhoT_centre}
\end{figure*}

Figure~\ref{fig:rhoT_centre} shows the thermal evolution at the centre of the cloud core for the grey (black) and the multigroup (red) 
simulations alongside results from other studies. The protostellar collapse occurs as follows:
\begin{itemize}
\item[$\bullet$] The cloud core first contracts under its own gravity and, as the cloud is optically thin, all the compression heating 
is radiated away; the cloud collapses isothermally.
\item[$\bullet$] As the density inside the core increases, the optical depth eventually surpasses unity and the cloud begins to retain 
the heat from compression. This is the formation of the first core ($M\sim2\times10^{-2}~\msun$, $R\sim10$ AU) and the core 
subsequently contracts adiabatically.
\item[$\bullet$] In the adiabatic phase, the temperatures are high enough for the rotational degrees of freedom of $\text{H}_{2}$ to 
be excited, and the effective adiabatic index $\gamma_{\text{eff}}$ is that of a diatomic gas ($=7/5$). The first core 
continues to contract and accrete infalling material.
\item[$\bullet$] When the temperature inside the first core reaches $\sim2000$ K, the molecules of $\text{H}_{2}$ begin to dissociate. 
This endothermic process constitutes an important energy sink which initiates the second phase of the collapse. During this phase, 
$\gamma_{\text{eff}}$ is usually approximated to about 1.1 \citepalias[see][]{masunaga2000}, although our results suggests that it is in 
fact somewhat higher.
\item[$\bullet$] Finally, once all the $\text{H}_{2}$ has been dissociated, the second collapse ends, the adiabatic regime is restored 
and the second core is formed ($M\sim10^{-3}~\msun$, $R\sim3\times10^{-3}~\text{AU}\simeq0.62~\text{R}_{\odot}$).
\end{itemize}

The thermal evolutions of the grey and multigroup simulations are very similar to each other. Small differences are visible at the time
of first hydrostatic core formation, with the multigroup simulation developing a core slightly earlier. The time of first core formation
corresponds to the time when the collapsing envelope becomes optically thick and the radiation can no longer escape from inside the system.
If the opacities are different in the grey and multigroup cases, one would expect the first cores to form at different times; in this
particular situation, it appears that the overal absorption of photons is more efficient in the multigroup case since the core becomes
adiabatic earlier.

The results are also in good agreement with the works of \citetalias{masunaga2000}, \citet{whitehouse2006}, \citetalias{stamatellos2007}
and \citetalias{tomida2013}. The centre of the core in the \citetalias{tomida2013} simulation during the first adiabatic contraction is
somewhat hotter than our results (and the other studies mentioned). This appears to be due to the use of a different EOS; in our simulation
$\gamma_{\text{eff}}$ starts to drop below 5/3 earlier than in the EOS used by \citetalias{tomida2013} (20 K compared to 100 K for
\citetalias{tomida2013}; see Fig.~\ref{fig:species} and Fig.~1 in \citetalias{tomida2013}). This is further illustrated by the orange
curve in Fig.~\ref{fig:rhoT_centre} which represents the thermal evolution of a simulation we ran with exactly the same setup as run 1 but
using an ideal gas EOS with a fixed $\gamma = 5/3$ instead of the \citetalias{saumon1995} EOS. We can clearly see that for densities below
$10^{-10}~\text{g~cm}^{-3}$, the \citetalias{tomida2013} behaves very much like an ideal gas with $\gamma = 5/3$. The EOS used by
\citet{whitehouse2006} seems to operate in a very similar manner, while the simulations of \citetalias{masunaga2000} and
\citetalias{stamatellos2007} follow our thermal track much more closely.

One of the main differences between the various EOS used by the different studies is the treatment of the different spin isomers of the
$\text{H}_{2}$ molecule in the low-to-moderate temperature regime. At the time of formation of the first core (and for a while later),
the gas is composed entirely of neutral $\text{H}_{2}$ and He, with $\text{H}_{2}$ being the dominant species. $\text{H}_{2}$ molecules
come in two forms corresponding to the two different spin configurations called para- (singlet state) and ortho- (triplet state) hydrogen. 
The \citetalias{saumon1995} EOS takes into account the symmetry of the nuclei wave-functions explicitly and makes no assumptions as to the
population ratios of the two species, which inherently implies thermodynamic equilibrium. However, the transition to ortho-para equilibrium
is known to be a lengthy process -- unless a magnetic catalyst (e.g. iron) is present in the medium -- so that \emph{at low temperatures}
($T \lesssim 300$ K) the population distribution of the two $\text{H}_{2}$ monomers is not the equilibrium value. Observations indeed
suggest that the real abundance ratio in molecular clouds and star forming regions is far from the thermal equilibrium value
\citep[see][for two recent examples]{pagani2011,dislaire2012}, even though large discrepancies (due to observational difficulties) between
the studies remain. For this reason, \citetalias{stamatellos2007} and \citetalias{tomida2013} have made the assumption that the ortho:para
abundance ratio remains frozen at its initial value of 3:1 (which reflects the statistical weight of each variety according to their spin
degeneracies), as ortho- and para-hydrogen form on the surface of dust grains. Using a fixed rather than an equilibrium ratio can
potentially have a significant impact on the early thermal evolution of the collapsing body (i.e. when temperatures remain below the spin
equilibrium temperature of $\sim 170$ K). On the other hand, \citet{flower1984} have shown that under typical molecular cloud conditions
($n \sim 100 - 1000~\text{cm}^{-3}$) the time to reach ortho:para equilibrium is of the order of 1 Myr. This is of course $5-10$ times
larger than the free-fall time of our system, but the latter is formed as a result of turbulence in the molecular cloud which spawns
over-densities that become gravitationally unstable. The collapse of a dense sphere (akin to our initial conditions with
$\rho \sim 10^{-19}~\text{g~cm}^{-3}$) will begin long after the formation of the molecular cloud which has a typical lifetime of
$\sim 10^{7}$ years and it is thus very possible that at the onset of the collapse, ortho:para equilibrium has already been reached. In
summary, it is not clear which ortho:para strategy (fixed ratio or equilibrium) is the most representative of the initial conditions of
star formation.

Different treatments of $\text{H}_{2}$ molecules will not affect the optically thin parts of the system where the gas temperature is
controlled by the radiation field (and the value of $\gamma_{\text{eff}}$ does not matter), but could explain why the simulation of
\citetalias{tomida2013} produces a first core which is hotter than our own for the same densities. A hotter first core can in turn have
an effect on the properties of the second core, since the $\text{H}_{2}$ dissociation temperature is reached earlier (in terms of central
density) and the second phase of the collapse will thus also end earlier (the amount of $\text{H}_{2}$ which has to be dissociated
remains the same). Consequently, the initial proto-stellar seed formed at the end of the second collapse will have a lower density and
its radius will probably be larger. As it turns out, \citetalias{tomida2013} find a second core which is also slightly more massive than
us (a factor of $\sim 2.5$), yielding a much ($\sim 10$ times) larger body (see the properties of the second core formed in our
simulations in section~\ref{sec:secondcollapseprofiles}).

It is however not possible to determine if the ortho/para $\text{H}_{2}$ treatment is the main contributor to the differences in
thermal evolutions. The only robust method would be to compute a new EOS table using the \citetalias{saumon1995} code, forcing the
ortho:para ratio to remain fixed at 3:1, but this procedure is rather complex and beyond the scope of this paper. Nevertherless, we can
speculate by looking more closely at the other studies. Indeed, \citetalias{stamatellos2007} have used the same assumption as
\citetalias{tomida2013} but their thermal evolution mirrors our curve. In contrast, the thermal evolution of the \citet{whitehouse2006}
study, which makes use of an equilibrium model from \citet{black1975} very similar to our own, tends to follow the \citetalias{tomida2013}
path. If the ortho:para ratio was the dominant factor, one would expect it to be the other way round (\citetalias{stamatellos2007} $\simeq$
\citetalias{tomida2013}; \citealt{whitehouse2006} $\simeq$ this work), which leads us to believe that the abundances of the $\text{H}_{2}$
flavours cannot alone be responsible for the discrepancies between the studies. 

Finally, we also note that even though they use the same EOS, the results of \citetalias{masunaga2000} show a second core forming later
than in our calculations; this is most probably due to a difference in opacities used.

The inset in Fig.~\ref{fig:rhoT_centre} shows a `bounce' in the thermal evolution. There is a time during the simulation when the core 
becomes thermally supported, stops contracting and begins to inflate. The density and gas temperature inside the adiabatic body 
decrease and the first core radius increases until the thermal pressure is no longer high enough to prevent contraction. The collapse 
then resumes and this time the infalling gas has enough momentum to drive the collapse past this point, to a state where gravity 
becomes once again dominant, enabling further contraction. This bounce was not seen by \citetalias{masunaga2000} and \citetalias{tomida2013} but 
it is visible (with a smaller amplitude) in Fig.~7 of \citetalias{stamatellos2007} (it is however not visible in our approximate 
reproduction of the \citetalias{stamatellos2007} data). A bounce is also visible in a very similar test in Fig.~14 of \citetalias{stamatellos2007}
but mention and/or discussion are absent from the text.

\begin{figure}[!ht]
\centering
\includegraphics[scale=0.415]{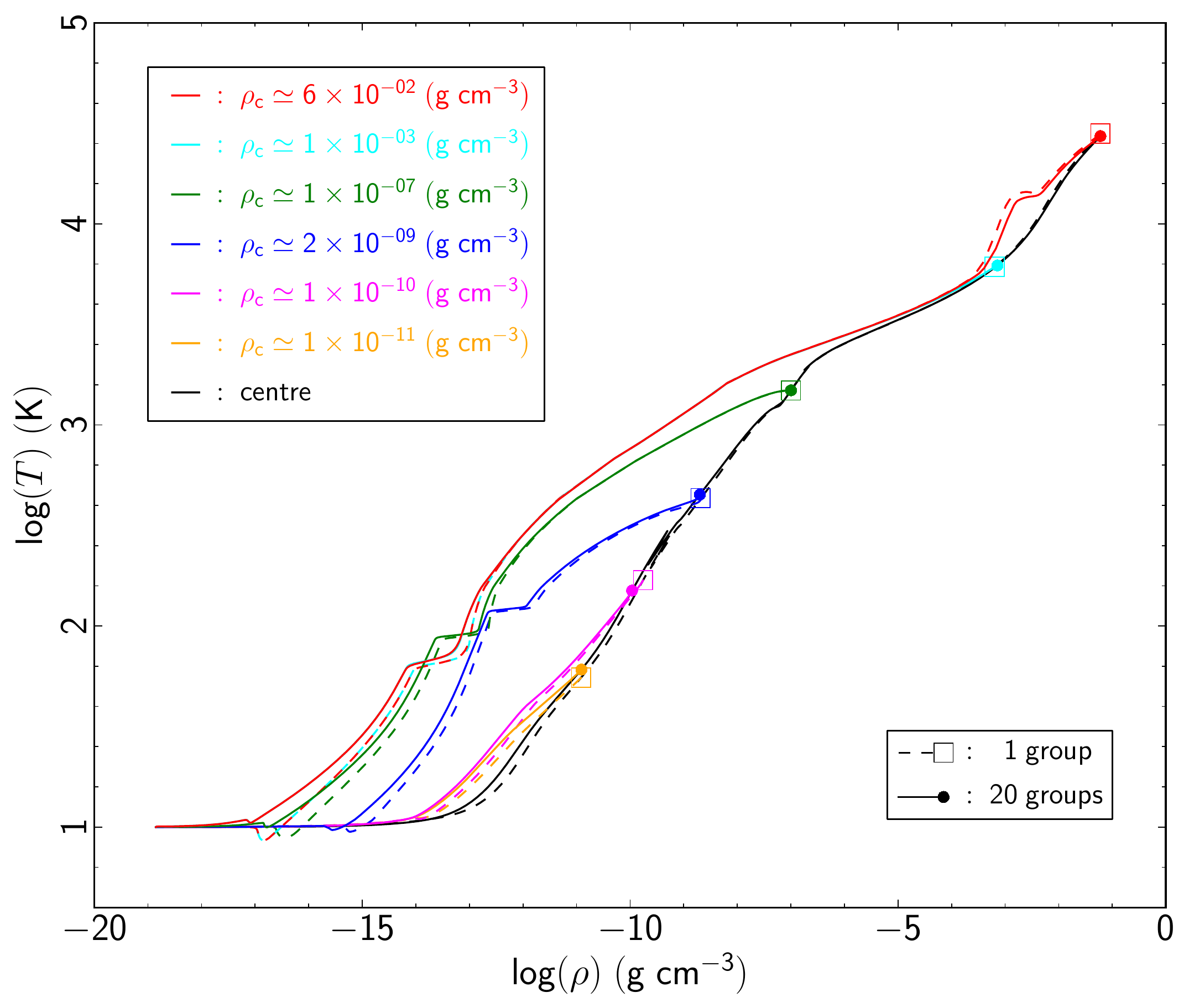}
\caption[Thermal evolution for 1 and 20 groups]{Thermal evolution for the grey (dashed) and 20-group (solid) simulations.}
\label{fig:rhoT_evol}
\end{figure}

Figure~\ref{fig:rhoT_evol} shows snapshots of the state of the gas in the system at six different epochs for the grey (dashed) and multigroup
(solid) simulations. The thermal evolution of the central fluid element from Fig.~\ref{fig:rhoT_centre} is also plotted for reference (black).
At early times ($\rho_{c} \leq 10^{-10}~\text{g~cm}^{-3}$), all the gas in the grid follows approximately the same thermal evolution as the
centre of the core. At later times, shock heating and absorption of radiation coming from the hot centre enable the outer layers of the system
to have much higher temperatures than the central point. We note here again that differences between the grey and multigroup simulations are
small. A displacement in the position of the first core accretion shock away from the centre is also visible on this plot; this is discussed
later in section~\ref{sec:firstcoreevol}.

\subsection{Radial profiles}\label{sec:secondcollapseprofiles}

\begin{figure*}[!ht]
\centering
\includegraphics[scale=0.38]{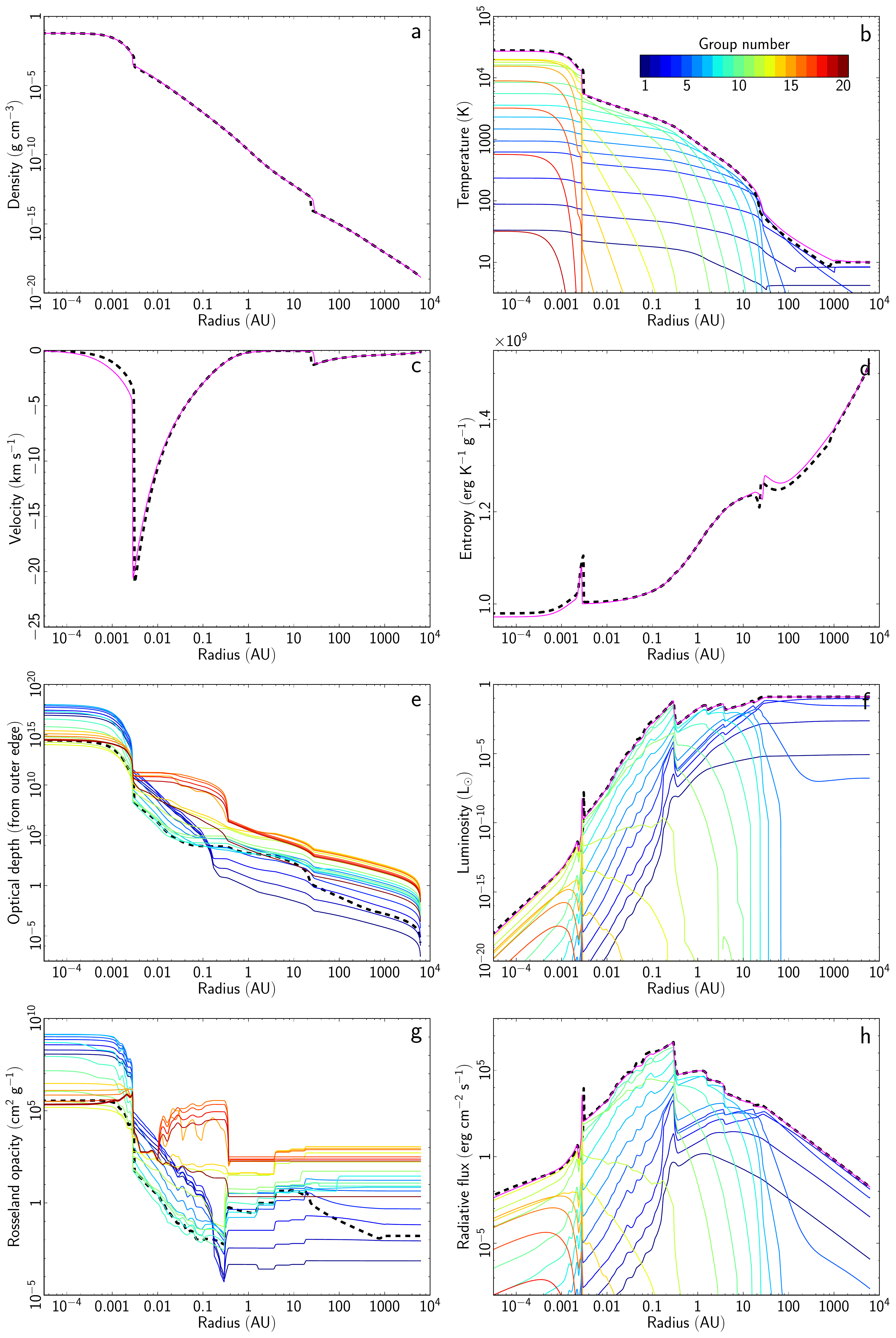}
\caption[Second collapse radial profiles for 1 \msun cloud with 20 groups]{Radial profiles of various properties during the collapse 
of a $1~\msun$ dense clump at a core central density $\rho_{c} = 6 \times 10^{-2}~\text{g~cm}^{-3}$, for the grey (black dashed) and 
multigroup (colours) models. For the multigroup run, the gas and total radiative (summed over all groups) quantities are plotted in 
magenta, while the other colours represent the individual groups (the colour-coding is the same as in Fig.~\ref{fig:kappanu}); see
legend in (b). From top left to bottom right: (a) density, (b) gas (magenta and black) and radiation (other colours) temperature, (c)
velocity, (d) entropy, (e) optical depth, (f) luminosity, (g) Rosseland average opacity and (h) radiative flux.}
\label{fig:second_collapse_profiles}
\end{figure*}

\begin{table}
\centering
\caption[Simulation parameters]{Initial conditions for the different simulations. Columns 2, 4 and 5 indicate the initial mass, radius
and temperature of the parent cloud, respectively. The third column specifies the number of frequency groups used in each run and the
sixth column lists the values for the thermal to gravitational potential energy ratio $\alpha$. Column 7 lists the free-fall
time of the initial could while the last column reports the time at the end of the simulation, when $\rho_{c} = 6 \times 10^{-2}~\text{g~cm}^{-3}$.}
\begin{tabular}{@{~}c@{~~~}c@{~~~}c@{~~~}c@{~~~}c@{~~~}c@{~~~}c@{~~~}c@{~}}
\hline
Run    & Mass of                    & Number of          & $R_{\text{init}}$         & $T_{\text{init}}$   & $\alpha$             & $t_{\text{ff}}$        & Time           \\
number & cloud                      & groups             & (AU)                      & (K)                 & ~                    & (Myr)                  & (Myr)          \\
\hline
 1     & \multirow{2}{*}{  1 \msun} &  1                 & \multirow{2}{*}{$10^{4}$} & \multirow{2}{*}{10} & \multirow{2}{*}{1.02}& \multirow{2}{*}{0.177} & 0.193          \\
 2     &                            & 20                 &                           &                     &                      &                        & 0.193          \\
\hline
 3     & \multirow{2}{*}{0.1 \msun} &  1                 & \multirow{2}{*}{$10^{3}$} & \multirow{2}{*}{10} & \multirow{2}{*}{1.02}& \multirow{2}{*}{0.018} & 0.021          \\
 4     &                            & 20                 &                           &                     &                      &                        & 0.022          \\
\hline
 5     & \multirow{2}{*}{ 10 \msun} &  1                 & \multirow{2}{*}{$10^{5}$} & \multirow{2}{*}{10} & \multirow{2}{*}{1.02}& \multirow{2}{*}{1.775} & 1.916          \\
 6     &                            & 20                 &                           &                     &                      &                        & 1.919          \\

\hline

 7     & \multirow{4}{*}{  1 \msun} & \multirow{4}{*}{1} & $5 \times 10^{3}$         & 10                  & 0.51                 & 0.063                  & 0.062          \\
 8     &                            &                    & $2 \times 10^{4}$         &  5                  & 1.02                 & 0.502                  & 0.551          \\
 9     &                            &                    & $10^{4}$                  &  5                  & 0.51                 & 0.177                  & 0.177          \\
10     &                            &                    & $5 \times 10^{3}$         & 20                  & 1.02                 & 0.063                  & 0.068          \\

\hline
\end{tabular}
\label{tab:simulation-params}
\end{table}

\begin{table*}
\centering
\caption[Core properties]{Summary of the first and second cores properties when $\rho_{c} = 6 \times 10^{-2}~\text{g~cm}^{-3}$ for the 
different simulations. The different columns list in order: core radius ($R$) and mass ($M$), mass accretion rate ($\dot{M}$), accretion
luminosity ($L_{\text{acc}}$) and radiated luminosity ($L_{\text{rad}}$) at the core border. $T_{\text{fc}}$ is the temperature at the
first core border, while $T_{c}$ is the temperature at the centre of the second core. $S_{c}$ represents the entropy at the centre of
the system when $\rho_{c} = 10^{-8}~\text{g~cm}^{-3}$ in the case of the first core, and at the end of the simulation for the second
core. The 9\textsuperscript{th} column represents the upstream Mach number of the flow at the accretion shock. $t_{\text{fc}}$ is the
lifetime of the first core, while the time given in the last column of the second core sub-table is the characteristic timescale of the
second core (see text).}
\begin{tabular}{@{~~~}c@{~~~}c@{~~~}c@{~~~}c@{~~~}c@{~~~}c@{~~~}c@{~~~}c@{~~~}c@{~~~}c@{~~~}c@{~~~}}
\hline
\multicolumn{11}{|c|}{First core} \\
\hline
Run    & $R$                   & $M$                   & $\dot{M}$             & $L_{\text{acc}}$      & $L_{\text{rad}}$      & $T_{\text{fc}}$      & $S_{c}$                             & $S_{N}$                             & Mach   & $t_{\text{fc}}$ \\ 
number & (AU)                  & ($\msun$)             & ($\msun/\text{yr}$)   & ($\lsun$)             & ($\lsun$)             & (K)                  & ($\text{erg~K}^{-1}~\text{g}^{-1}$) & ($\text{erg~K}^{-1}~\text{g}^{-1}$) & number & (yr)            \\
\hline
 1     & 24.1                  & $4.34 \times 10^{-2}$ & $3.53 \times 10^{-5}$ & $8.73 \times 10^{-3}$ & $1.16 \times 10^{-1}$ &   64                 & $9.79 \times 10^{8}$                & $2.26 \times 10^{9}$                & 2.51   &  886            \\
 2     & 28.2                  & $4.71 \times 10^{-2}$ & $3.39 \times 10^{-5}$ & $7.63 \times 10^{-3}$ & $1.09 \times 10^{-1}$ &   66                 & $9.83 \times 10^{8}$                & $2.27 \times 10^{9}$                & 2.27   &  982            \\
\hline
 3     & 33.3                  & $4.97 \times 10^{-2}$ & $1.49 \times 10^{-5}$ & $3.04 \times 10^{-3}$ & $1.22 \times 10^{-1}$ &   44                 & $9.72 \times 10^{8}$                & $2.25 \times 10^{9}$                & 2.75   & 2121            \\
 4     & 39.6                  & $5.36 \times 10^{-2}$ & $1.34 \times 10^{-5}$ & $2.49 \times 10^{-3}$ & $1.12 \times 10^{-1}$ &   42                 & $9.75 \times 10^{8}$                & $2.25 \times 10^{9}$                & 2.57   & 2424            \\
\hline
 5     & 20.5                  & $4.04 \times 10^{-2}$ & $4.93 \times 10^{-5}$ & $1.33 \times 10^{-2}$ & $1.15 \times 10^{-1}$ &   84                 & $9.83 \times 10^{8}$                & $2.27 \times 10^{9}$                & 2.18   &  614            \\
 6     & 23.0                  & $4.28 \times 10^{-2}$ & $4.88 \times 10^{-5}$ & $1.23 \times 10^{-2}$ & $1.06 \times 10^{-1}$ &   90                 & $9.86 \times 10^{8}$                & $2.28 \times 10^{9}$                & 1.93   &  644            \\

\hline

 7     & 5.99                  & $2.36 \times 10^{-2}$ & $1.15 \times 10^{-4}$ & $6.23 \times 10^{-2}$ & $5.65 \times 10^{-2}$ &  320                 & $9.94 \times 10^{8}$                & $2.30 \times 10^{9}$                & 1.25   &  148            \\
 8     & 35.0                  & $5.09 \times 10^{-2}$ & $1.35 \times 10^{-4}$ & $2.69 \times 10^{-3}$ & $1.36 \times 10^{-1}$ &   44                 & $9.68 \times 10^{8}$                & $2.24 \times 10^{9}$                & 2.76   & 2638            \\
 9     & 21.2                  & $4.08 \times 10^{-2}$ & $4.40 \times 10^{-5}$ & $1.16 \times 10^{-2}$ & $1.13 \times 10^{-1}$ &   77                 & $9.78 \times 10^{8}$                & $2.26 \times 10^{9}$                & 2.37   &  763            \\
10     & 6.26                  & $2.35 \times 10^{-2}$ & $9.97 \times 10^{-5}$ & $5.10 \times 10^{-2}$ & $5.49 \times 10^{-2}$ &  304                 & $1.00 \times 10^{9}$                & $2.32 \times 10^{9}$                & 1.20   &  131            \\

\hline
\multicolumn{11}{c}{~}\\
\hline

\multicolumn{11}{|c|}{Second core}\\
\hline
Run    & $R$                   & $M$                   & $\dot{M}$             & $L_{\text{acc}}$      & $L_{\text{rad}}$      & $T_{c}$              & $S_{c}$                             & $S_{N}$                             & Mach   & $\mathcal{T}_{\text{sc}}$ \\ 
number & (AU)                  & ($\msun$)             & ($\msun/\text{yr}$)   & ($\lsun$)             & ($\lsun$)             & (K)                  & ($\text{erg~K}^{-1}~\text{g}^{-1}$) & ($\text{erg~K}^{-1}~\text{g}^{-1}$) & number & (Gyr)                     \\
\hline
 1     & $3.07 \times 10^{-3}$ & $1.34 \times 10^{-3}$ & $2.01 \times 10^{-1}$ & $1.24 \times 10^{4}$  & $1.61 \times 10^{-8}$ & $2.81 \times 10^{4}$ & $9.81 \times 10^{8}$                & $1.27 \times 10^{9}$                & 3.65   & 2.27                      \\
 2     & $2.81 \times 10^{-3}$ & $1.23 \times 10^{-3}$ & $2.37 \times 10^{-1}$ & $1.47 \times 10^{4}$  & $5.63 \times 10^{-9}$ & $2.73 \times 10^{4}$ & $9.73 \times 10^{8}$                & $1.26 \times 10^{9}$                & 3.43   & 3.63                      \\
\hline
 3     & $3.07 \times 10^{-3}$ & $1.34 \times 10^{-3}$ & $2.02 \times 10^{-1}$ & $1.25 \times 10^{4}$  & $1.52 \times 10^{-8}$ & $2.82 \times 10^{4}$ & $9.81 \times 10^{8}$                & $1.27 \times 10^{9}$                & 3.65   & 1.34                      \\
 4     & $2.85 \times 10^{-3}$ & $1.24 \times 10^{-3}$ & $2.26 \times 10^{-1}$ & $1.39 \times 10^{4}$  & $6.21 \times 10^{-9}$ & $2.75 \times 10^{4}$ & $9.74 \times 10^{8}$                & $1.26 \times 10^{9}$                & 3.49   & 2.88                      \\
\hline
 5     & $3.09 \times 10^{-3}$ & $1.34 \times 10^{-3}$ & $2.01 \times 10^{-1}$ & $1.24 \times 10^{4}$  & $1.71 \times 10^{-8}$ & $2.81 \times 10^{4}$ & $9.81 \times 10^{8}$                & $1.27 \times 10^{9}$                & 3.66   & 0.89                      \\
 6     & $2.89 \times 10^{-3}$ & $1.25 \times 10^{-3}$ & $2.26 \times 10^{-1}$ & $1.40 \times 10^{4}$  & $8.96 \times 10^{-9}$ & $2.74 \times 10^{4}$ & $9.74 \times 10^{8}$                & $1.26 \times 10^{9}$                & 3.48   & 1.54                      \\

\hline

 7     & $3.25 \times 10^{-3}$ & $1.53 \times 10^{-3}$ & $2.14 \times 10^{-1}$ & $1.43 \times 10^{4}$  & $2.15 \times 10^{-8}$ & $3.08 \times 10^{4}$ & $1.01 \times 10^{9}$                & $1.29 \times 10^{9}$                & 3.87   & 0.80                      \\
 8     & $3.18 \times 10^{-3}$ & $1.38 \times 10^{-3}$ & $1.94 \times 10^{-1}$ & $1.20 \times 10^{4}$  & $1.97 \times 10^{-8}$ & $2.82 \times 10^{4}$ & $9.82 \times 10^{8}$                & $1.27 \times 10^{9}$                & 3.70   & 0.75                      \\
 9     & $3.10 \times 10^{-3}$ & $1.35 \times 10^{-3}$ & $1.98 \times 10^{-1}$ & $1.22 \times 10^{4}$  & $1.53 \times 10^{-8}$ & $2.81 \times 10^{4}$ & $9.81 \times 10^{8}$                & $1.27 \times 10^{9}$                & 3.67   & 0.96                      \\
10     & $3.27 \times 10^{-3}$ & $1.55 \times 10^{-3}$ & $2.14 \times 10^{-1}$ & $1.44 \times 10^{4}$  & $2.06 \times 10^{-8}$ & $3.12 \times 10^{4}$ & $1.01 \times 10^{9}$                & $1.30 \times 10^{9}$                & 3.91   & 0.85                      \\
                           
\hline
\end{tabular}
\label{tab:coreprops}
\end{table*}

Figure~\ref{fig:second_collapse_profiles} shows the radial profiles of the density, temperature, velocity, entropy, optical depth, 
luminosity, opacity and radiative flux for the grey (black dashed line) and multigroup simulations (colours) for a central density
$\rho_{c} = 6 \times 10^{-2}~\text{g~cm}^{-3}$. The first and second core borders are visible at $\sim 30$ AU and $3\times10^{-3}$
AU respectively, this being most clear in the density (a) and velocity (c) panels. The temperature plot (b) reveals that the first
core accretion shock is supercritical \citep[pre- and post-shock temperatures are equal, see discussion in][]{commercon2011} while
the shock at the second core border is subcritical \citepalias[the simulations of][also show this]{tomida2013}.
Table~\ref{tab:coreprops} lists the main properties of the first and second cores; these are the core radius ($R$) and mass ($M$),
the mass accretion rate at the core border ($\dot{M}$), the accretion luminosity ($L_{\text{acc}}$), the total radiated luminosity
($L_{\text{rad}}$), the temperature at the first core border ($T_{\text{fc}}$) and at the centre of the second core ($T_{c}$), the
entropy at the centre of the system ($S_{c}$), the accretion shock Mach number, the first core lifetime ($t_{\text{fc}}$) and the
time in the simulation when the central density has reached $6 \times 10^{-2}~\text{g~cm}^{-3}$. Further details on the derivation
of these quantities can be found in \citetalias{vaytet2012}.

Compared to the first collapse simulations in \citetalias{vaytet2012}, the first core has now grown both in size and mass, from 7
to about 30 AU and from $2\times10^{-2}~\msun$ to $\sim 4 \times 10^{-2}~\msun$. We also notice that the temperatures at the first
core border is half the value reported in \citetalias{vaytet2012}, probably because of its increased size and the use of a slightly
different EOS. The first core lifetime $t_{\text{fc}}$ is defined as the time elapsed between the formation of the first core
(chosen as the time when $\rho_{c} > 3 \times 10^{-10}~\text{g~cm}^{-3}$) and the beginning of the second collapse (when the central
temperature exceeds 2000 K). The subsequent formation of the second core, after which the spectral properties of the collapsing
system change dramatically \citetalias[see][for example]{masunaga2000}, is almost instantaneous (see sections
\ref{sec:secondcoreformation} and \ref{sec:firstcoreevol}). Our simulation yields a lifetime of $\sim 1000$ years, which is a
relatively short time for a chance to observe a first core in its formation stage. This value is of course a lower limit because
no support from rotation is present in our spherically symmetric model.

The second core is very compact, measuring only $3\times10^{-3}$ AU in size for a mass of $10^{-3}~\msun$. The accretion luminosity,
which is defined as
\begin{equation}\label{eq:lacc}
L_{\text{acc}} = \frac{GM\dot{M}}{R} ~,
\end{equation}
is an estimate of the luminosity at the accretion shock assuming that all the infalling kinetic energy is transformed into radiation.
At the second core border, $L_{\text{acc}}$ greatly outweighs the total radiative luminosity (by about 12 orders of magnitude) which,
together with the fact that the shock is subcritical, shows that all the accretion energy is transfered to the second core, none of
it is radiatied away; the accretion shock is almost completely adiabatic. This strongly differs from what happens at the first core
border where the vast majority of accretion energy is transformed into radiation \citepalias[see][]{vaytet2012}. Note that the
situation could be different in 3D simulations where a significant fraction of the accretion energy may be transported away by
outflows. We also remark that the mass accretion rates at the second core border are colossal ($0.2~\msun/\text{yr}$); the second
core's mass is growing very rapidly.

\citet{stahler1980} predicted that whether the second core accretion shock would be sub- or supercritical would depend on the
magnitude of a dimensionless parameter $\kappa \rho R$, computed just ahead of the shock ($R$ is the shock radius). For
$\kappa \rho R \ll 1$ the should would be subcritical whereas in the case of $\kappa \rho R \gg 1$ the shock would be supercritical.
This parameter does not make any sense to us because what determines whether a shock is in the sub- or supercritical regime is the
optical depth of the gas downstream and more importantly upstream of the shock \citep[see][pages 296+ for a discussion]{drake2006}.
Here, $\kappa \rho R$ is only a simple estimate of the optical thickness of the gas downstream of the shock (assuming $\rho$ and
$\kappa$ to be constants inside the core), and the upstream condition is ignored altogether. In addition, if we consider their
subcritical scenario, they estimate that about 3/4 of the accreted energy is transfered to the core; as mentioned above we find it
to be $\sim 1$.

There are two peaks in radiative flux; the first one around 0.3 AU originates from the region where the opacity has fallen sharply due 
to the destruction of dust grains when the temperature exceeds 1500 K (see gaps in Figs~\ref{fig:second_collapse_profiles}g and 
\ref{fig:kappagrey}). This represents the dust border \citep[see][for example]{stahler1980}, but only the radiative flux is strongly 
affected at that location, the hydrodynamical quantities seem relatively insensitive to the presence of this border, except for a 
small kink visible in the temperature profile. The second burst in radiative flux occurs at the second core border where the sharp 
jump in gas and radiative temperature provoke a rise in radiative flux. The high optical depth at that radius means that the flux is 
very rapidly absorbed and the flux burst only appears as a sharp spike in the profiles.

Surprisingly, the grey and multigroup simulations yield very similar results. The curves overlap in most places, only the core 
borders are at slighty different locations and the multigroup gas temperature is above its grey counterpart between 20 and 1000 AU.
Differences of $\sim 10-20$\% in the sizes and masses of the first and second cores are reported in Table~\ref{tab:coreprops}, which
are significant from a theoretical point of view but are in no way detectable through obervations. The second collapse is a very short
and violent and dynamic event in the lifetime of proto-stars, and the details of the radiation transport may not have time to affect
the system dynamics significantly. However, multi-frequency radiative transfer could have a greater effect on the long-term evolution
of the proto-star. One can estimate the characteristic timescale of the second core by integrating the total energy inside the core
and computing how long it will take for the core to radiate all of its energy with the current luminosity at the core border, that is
\begin{equation}\label{equ:timescale}
\mathcal{T}_{\text{sc}} = \frac{E_{\text{tot}}}{L_{\text{rad}}}
\end{equation}
where
\begin{equation}\label{equ:total_energy}
E_{\text{tot}} = \int_{0}^{R} \left(\frac{1}{2} \rho u^{2} + \rho e + E_{r} - \frac{4 \pi G \rho r^{3}}{3 r} \right) 4\pi r^{2} dr
\end{equation}
where $u$ is the gas velocity, $e$ the specific internal energy, $E_{r}$ the radiative energy and $G$ the gravitational constant.
The characteristic timescales are listed in the last column of the second core sub-table in Table~\ref{tab:coreprops}. We can see that
even though differences in core mass, radius and luminosity between grey and multigroup simulations are small, they can lead to
considerable variations in characteristic timescales. The subsequent evolution of the proto-star is of course very complex, with
continued accretion and the ignition of thermonuclear reactions, and we are obviously not making any strong claims with our simple
estimate, but merely suggesting that multi-frequency effects might be more significant in the long run than what is visible here.

\subsection{The second core formation}\label{sec:secondcoreformation}

\begin{figure*}[!ht]
\centering
\includegraphics[scale=0.40]{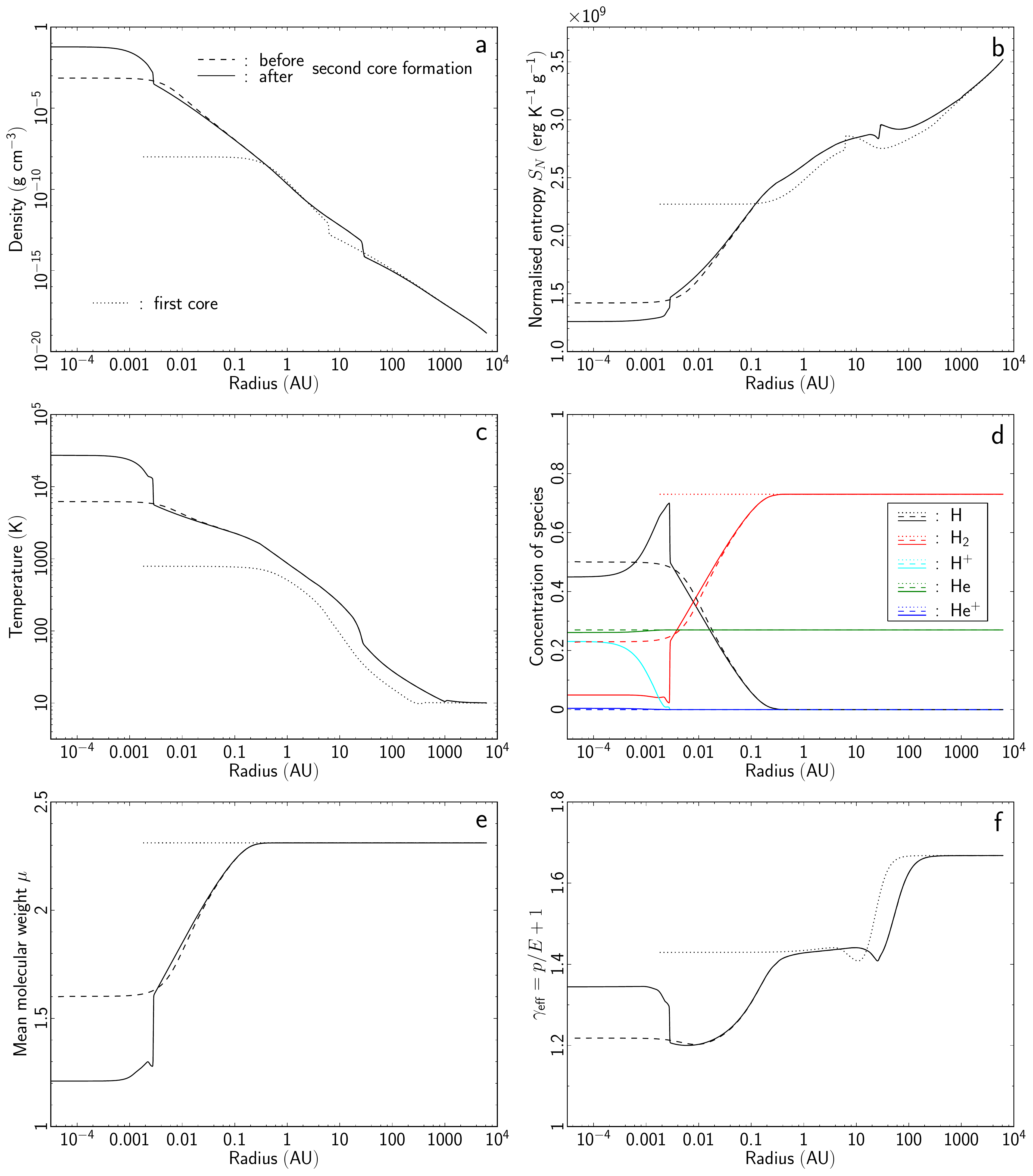}
\caption[Second core formation]{Radial profiles of the (a) gas density, (b) normalised entropy $S_{N}$ (see text), (c) temperature,
(d) species concentrations, (e) mean molecular weight $\mu$ and (f) effective ratio of specific heats $\gamma_{\text{eff}}$. In all
panels, the dotted line describes the first core profiles while the dashed and solid lines represent the quantities just before and
after the formation of the second core, respectively.}
\label{fig:species}
\end{figure*}

Figure~\ref{fig:species} illustrates what happens at the time of second core formation; the different panels show the radial profiles
of the gas density (a), normalised entropy $S_{N}$ (b), temperature (c), species mass concentrations (d), mean molecular weight $\mu$ (e)
and effective ratio of specific heats $\gamma_{\text{eff}}$ (f). $S_{N}$ is defined as the entropy per free particle multiplied by
$\rho/m_{H}$ which is simply
\begin{equation}\label{normalised_entropy}
S_{N} = \frac{S}{\sum_{i}X_{i}/A_{i}}
\end{equation}
where $i$ is summed over all the species, $X$ is the species mass concentration and $A$ their atomic number. The dotted line describes
the first core profiles while the dashed and solid curves represent the state of the system just before and after the formation of
the second core, respectively.

The first core profiles in panels (d) and (e) show that it is constituted entirely of $\text{H}_{2}$ and neutral He with a constant
mean molecular weight of 2.31. The dashed lines represent the profiles of the system when the second collapse is already well
underway but the second core has not yet materialised. Between radii of 0.5 and $10^{4}$ AU, the hydrogen and helium concentrations
as well as the mean molecular weight remain constant. Below 0.5 AU, the dissociation of $\text{H}_{2}$ starts to take place as the
gas temperature exceeds 2000 K, and the fraction of atomic hydrogen increases, which consequently causes $\mu$ to decrease. The
molecular and atomic hydrogen concentrations exhibit symmetric profiles, as a rise in one is compensated by a fall in the other.
They both reach a plateau below $5\times 10^{-3}$ AU, as is the case for the gas density, temperature and pressure.

The formation of the second core is very abrupt; the time between the two outputs (dashed and solid) is approximately two days. The
second core is formed as the dissociation of $\text{H}_{2}$ begins to shut down. In the centre, as most the $\text{H}_{2}$ is
destroyed, the energy sink provided by the dissociation no longer operates, and this starts to prevent further collapse. As the outer
infalling material smashes into the gas at the centre which is no longer collapsing, a strong hydrodynamical shock is created at the
border. The temperature and density of the accreted material increase sharply as they flow through the shock. We can see that
downstream from the shock, almost all the $\text{H}_{2}$ has been dissociated, a third of the atomic H has been ionised and there is
also a very small amount of He that gets ionised because of the high gas temperature.

The entropy of the gas at the centre of the system $S_{c}$ listed in Table~\ref{tab:coreprops} is identical for the first and second
cores; i.e. the first core sets the properties of the system. However, the normalised entropy $S_{N}$ in Fig.~\ref{fig:species}b shows
that the entropy per free particle decreases significantly between the first and second cores, due to the dissociation of $\text{H}_{2}$
which increases the number of particles. Further decrease is seen between the dashed and solid profiles because of the additional
dissociation taking place inside the second core.

Figure~\ref{fig:species}f shows the effective ratio of specific heats $\gamma_{\text{eff}} = p/e + 1$ where $p$ is the gas pressure
and $e$ the gas internal gas energy. It shows that the initial cloud starts out as a monatomic ideal gas ($\gamma_{\text{eff}} = 5/3$)
and transitiones to a diatomic gas ($\gamma_{\text{eff}} \simeq 7/5$) as the temperature exceeds 20 K where the rotational degrees of
freedom of the $\text{H}_{2}$ molecules begin to be excited. Inside the first core (between 0.5 and 20 AU), the gas is akin to a
diatomic adiabatic polytrope ($\gamma_{\text{eff}} \simeq 7/5$). Then, during the second collapse, the effective $\gamma_{\text{eff}}$
drops as low as 1.2 amid a phase transition from $\text{H}_{2}$ to H. Finally, inside the newly formed second core, the gas essentially
mono-atomic, and we expect $\gamma_{\text{eff}}$ to return to 5/3. However, there is $\sim 5$\% of $\text{H}_{2}$ remaining inside the
core and a small amount of dissociation is still operating, which lowers $\gamma_{\text{eff}}$. In addition, at such high temperatures
and densities, correlation effects start to become noticeable which also alter the value of $\gamma_{\text{eff}}$
\citep[see][]{saumon1992}.

\subsection{Varying the initial parameters}\label{sec:var_params}

\begin{figure*}[!ht]
\centering
\includegraphics[scale=0.48]{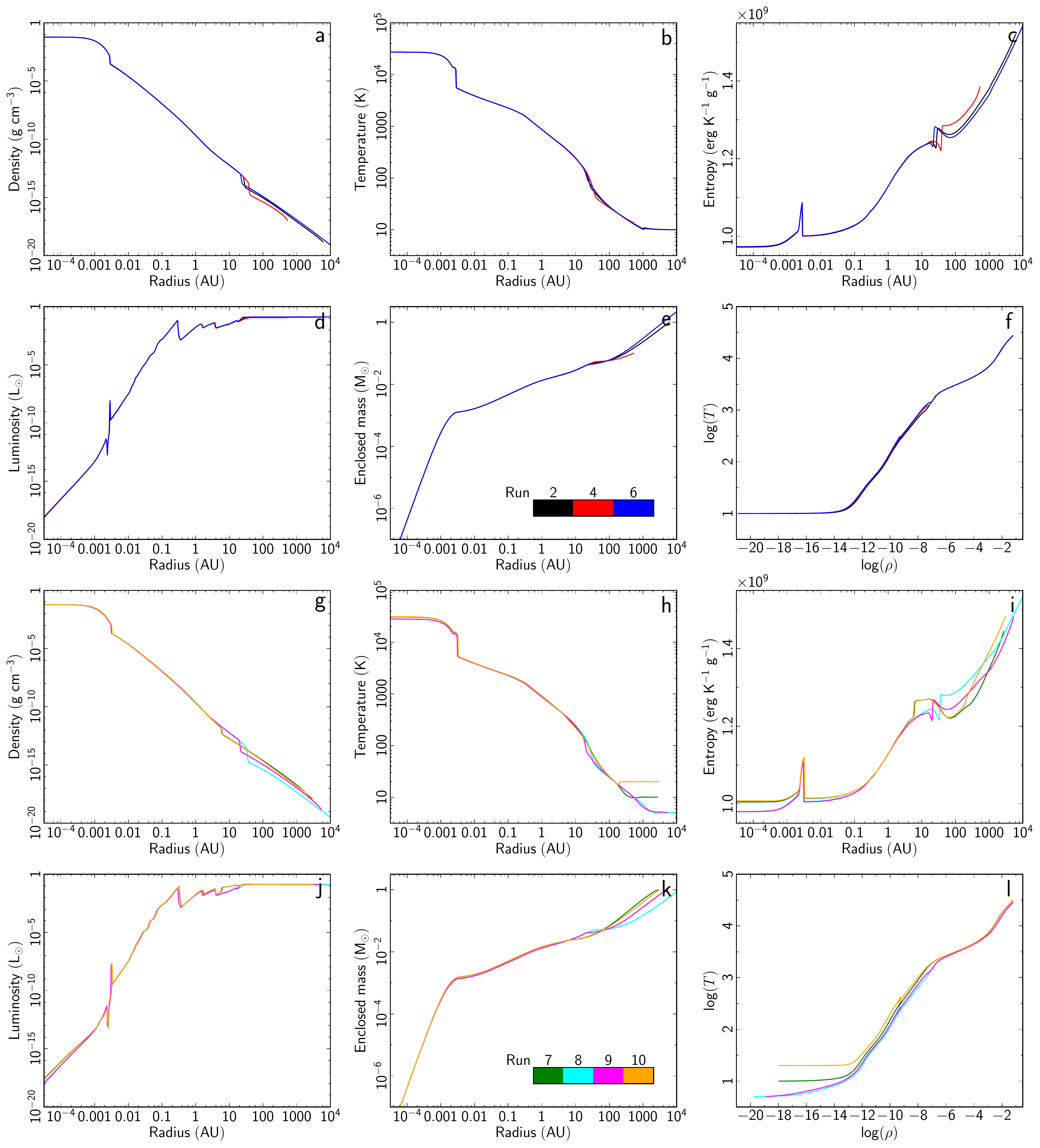}
\caption[First and second core radial profiles for different parameters]{Comparison of the radial profiles of collapse simulations
at a central density of $\rho_{c} = 6 \times 10^{-2}~\text{g~cm}^{-3}$ using various initial parameters (see Table~\ref{tab:coreprops}
for details). The different panels display the following as a function of raidus: (a) and (g) density, (b) and (g) gas temperature,
(c) and (i) entropy, (d) and (j) radiative luminosity, (e) and (k) enclosed mass. Panels (f) and (l) display the thermal evolution at
the centre of the grid. Panels (e) and (k) show the colour legend.}
\label{fig:all_collapses}
\end{figure*}

In order to check the universality/robustness of the results above, we also performed simulations of the collapse of a 0.1~\msun and
a 10~\msun cloud using 1 and 20 frequency groups. The initial setups were identical to that of \citetalias{vaytet2012} in that the
thermal to gravitational energy ratio was kept constant (see Table~\ref{tab:coreprops} for details). The results are shown in 
Fig.~\ref{fig:all_collapses} (panels a to f), along with the previous results from the 1~\msun case.

As in \citetalias{vaytet2012}, the results are strikingly similar to the 1~\msun case, for both 0.1~\msun and a 10~\msun clouds. The 
properties of the first and second cores for the new cloud masses are listed in Table~\ref{tab:coreprops}, which underlines the point 
further. The masses, temperatures and sizes of both cores seem invariant of the initial cloud mass, with the second core showing the 
highest convergence between models. Second core radii differ by less than 3\% while the masses exhibit variations of only 2\%. The
sizes and masses of the second cores also agree fairly well with the analytical estimates of \citet{baraffe2012}. All the simulations
begin with the same gravitational to thermal energy ratio, and it is thus not so surprising to see such a tight concordance of results.
We also add that in all cases, differences between grey and multigroup simulations remain very small (for the sake of clarity this is
not shown in the figures).

A further four simulations were also run, this time changing the initial gravitational to thermal energy ratio and the temperature
of the parent gas cloud (see Table~\ref{tab:coreprops} for details). Runs 7 and 8 were run with a parent cloud half and double the
size, respectively (the gas in run 8 was colder so that the cloud would collapse). The initial temperature of the gas in runs 9 and
10 was 5 and 20 K, respectively (the size of the cloud in run 10 was halved to overcome the stronger thermal support provided by
the hotter gas). The radial profiles are displayed in Fig.~\ref{fig:all_collapses} (panels g to l).

The different simulations yield similar results, and the sizes and masses (listed in Table~\ref{tab:coreprops}) of the second cores
further confirm their insensitivity and ignorance of the initial conditions. The size of the first core varies from 6 AU for runs 7
and 10 to about 20-30 AU for the other runs. 6 AU is the size the first core has at its time of formation \citepalias[see][]{vaytet2012}
and we explore the origin of the expansion of the first core in the next section.

\subsection{The evolution of the first core}\label{sec:firstcoreevol}

\begin{figure*}[!ht]
\centering
\includegraphics[scale=0.6]{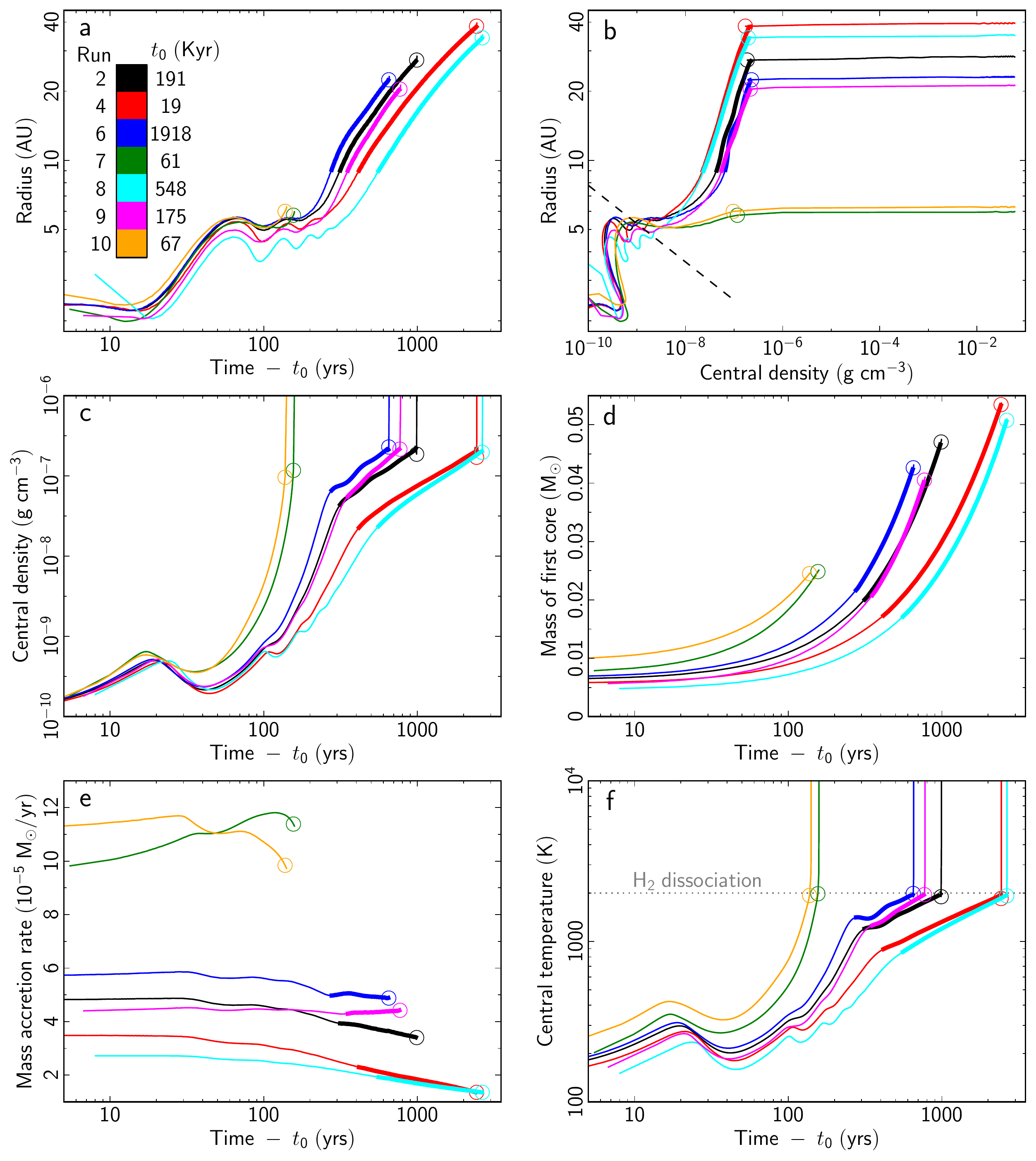}
\caption[First core mass and radius]{Evolution of the first core for runs 2, 4, 6, 7, 8, 9 and 10 (see colour legend in top left
panel). (a) Core radius as a function of time. (b) Core radius as a function of central density. The dashed line represents the
\citet{masunaga1998} estimate. (c) Central density as a function of time. (d) First core mass as a function of time. (e) Mass
accretion rate at the core border as a function of time. (f) Central temperature as a function of time. $t_{0}$ represents the
time when $\rho_{c} \ge 10^{-10}~\text{g~cm}^{-3}$, taken as the time of first core formation; $t_{0}$ is listed for each run
next to the colour key in panel (a). In all panels, the circles mark the onset of the second collapse and the bold lines trace
the `transition' region when the central density and temperature increase slower than during the rest of the simulation (see
text).}
\label{fig:core_mass_radius}
\end{figure*}

As mentioned in sections~\ref{sec:secondcollapsephases} and \ref{sec:var_params}, the first core border is not stationary throughout
the simulations, and this is further illustrated by the temporal evolutions plotted in Fig.~\ref{fig:core_mass_radius}. The timeframe
shown begins when the central density reaches $10^{-10}~\text{g~cm}^{-3}$ (taken as $t_{0}$). At early times ($t < 200$ yr), all the
simulations show one or several `bounces' (panels a, c and f) during which the first core is momentarily thermally supported and
oscillates between thermal and gravitational pressure\footnote{Note that a drop in central density or temperature coincides with an
increase in core radius, as expected for a gas sphere (almost) in hydrostatic equilibrium.}. This was already visible in
Fig.~\ref{fig:rhoT_centre}, and \citetalias{tomida2013} also observed similar `bounces'. Alongside is plotted in panel (b) the
evolution of the core radius as a function of the central density, which again shows oscillations in the lower left corner.

In the case of run 2 (black), after a small period of time during which it remains approximately constant
($125~\text{yr} < t < 225~\text{yr}$), the first core radius enters a phase of steady increase. Panel (e) shows that the core has
an almost constant accretion rate during that phase and is continuously growing in mass, while panel (c) reveals that the central
density, for most of the core's lifetime ($300~\text{yr} < t < 950~\text{yr}$), does not increase as strongly as it did at earlier
times; this `transition' period has been highlighted in bold on the figure. A more massive core with the same density can only be
larger, which explains the increase in core border radius. A second possible contribution to the inflation of the first core comes
from the radiation from the hot centre of the core which heats the gas in the outer layers (see Fig.~\ref{fig:rhoT_evol}), causing
it to expand. This increase in size was also found by \citet{schonke2011}, but is in disagreement with the analytical analysis of
\citet{masunaga1998} who predicted that the first core radius would decrease in time (see dashed line in panel b). We believe that
the discrepancy's origin lies with \citeauthor{masunaga1998}'s assumption that the first core is isentropic, which is not strictly
the case, as seen in Fig.~\ref{fig:second_collapse_profiles}d. Radiative transfer is capable of re-distributing entropy outwards,
which invalidates \citeauthor{masunaga1998}'s assumptions. It is a shame that they do not show in their \citeyear{masunaga2000}
paper the results they obtained when running new simulations with the same EOS \citepalias{saumon1995} as we have used in this work,
which would have allowed us to conduct a better comparison.

All runs but two, namely 7 and 10, follow a similar evolution pattern, with a lengthy `transition' phase (bold) of slowly
increasing central density and temperature. As for runs 7 (green) and 10 (orange), the behaviour is markedly different; the core
radius remains approximately constant during the entire core lifetime. The parent cloud in these two runs is half the size of that
of run 2 and is therefore more unstable (much smaller free-fall time; see Table~\ref{tab:simulation-params}), i.e. it collapses
faster (in simple terms, dividing the radius of the parent cloud by its free-fall time gives a dimensional estimate of the infall
velocity which scales with $r^{-1/2}$ for a given cloud mass). The higher infall velocity yields a larger mass accretion rate at
the core border (see panel d). The increased effects of gravity, on a core fast becoming more massive, boost contraction further
and in the process enhance heating at the centre. The 2000 K mark is reached earlier (see green and orange curves in panel f) and
the second collapse begins before the first core has had time to grow (no `transition' period is visible in the time profiles).
In addition, radiative heating of the outer layers of the core is again present in runs 7 and 10, but it seems incapable of
driving the core expansion against the strong ram pressure applied by the infalling matter at the core border. Finally, even
though the results for core radius as a function of central density are very different from the run 2 results, they are still not
in agreement with \citeauthor{masunaga1998}'s analytical estimate (see panel b). The free-fall time (or by extension the initial
cloud density) appears to be the dominant factor in setting the subsequent size of the first core.

Figure~\ref{fig:core_mass_radius} also shows again how sudden the second collapse phase is, with the central density
shooting up almost instantaneously (on the plotted timescale) in panel (a).

\subsection{Impact of the mass of Larson's second core for early protostar evolution}\label{sec:second_core_impact}

\citet{baraffe2009} and \citet{baraffe2010} showed that episodic accretion on a newborn protostar provides a plausible explanation
for the observed luminosity spread in young stellar clusters and star forming regions without invoking any age spread and can also
explain observed unexpectedly high depletion levels of lithium in some young objects. This scenario was questioned by
\citet{hosokawa2011} who argued that the scenario could not hold in the lower ($T_{\text{eff}} \lesssim 3500$ K) part of the
Hertzsprung-Russell diagram. The issue has been addressed in detail in \citet{baraffe2012}, where the authors have shown that the
only reason why \citet{hosokawa2011} could not reproduce the observed luminosity spread in the aforementioned domain stems from their
assumed value for the second Larson core mass (i.e. the protostar initial mass), namely $10~\text{M}_{\text{jup}}$, a value more
representative of the \emph{first} Larson core \citepalias[see][for example]{vaytet2012}. Using smaller values, in particular
$1~\text{M}_{\text{jup}}$ or so, \citet{baraffe2012} adequately reproduced the observed spread within the very same episodic accretion
scenario. \citet{baraffe2012} thus confirmed a unified picture for early evolution of accreting protostars and concluded that the
controversy raised by \citet{hosokawa2011} should be closed, \emph{except} if it was shown unambiguously that the initial protostar/BD
mass could not be smaller than $10~\text{M}_{\text{jup}}$. The present calculations, pointing to a ``universality'' of the second
core mass of about $1~\text{M}_{\text{jup}}$ thus agree with the analytical estimate of \citet{baraffe2012} and confirm their
conclusions.

\section{Conclusions}\label{sec:conclusions}

We have performed multigroup RHD simulations of the gravitational collapse of a 1~\msun cold dense cloud core up to the formation of 
the second Larson core, reaching a central density of $\rho_{c} = 6 \times 10^{-2}~\text{g~cm}^{-3}$. Twenty groups were used to 
sample the opacities in the frequency domain and the results were compared to a grey simulation. Only small differences were found 
between the two runs, with no major structural or evolutionary changes. The main properties of the resulting first and second cores 
formed in the centre of the grid such as their mass and size exhibited differences of $\sim 10-20$\%, which is substantial from a 
theoretical standpoint, but appears relatively unsignificant/undetectable in observational studies (note that the gas entropy inside
the cores were almost identical between the two simulations).

Nevertherless, we found that following its formation, the first core continues to accrete envelope material, steadily growing in mass
and size. By the time the second core is formed, its radius has increased by a factor of 5 to 6 \citep[a result in disagreement with
the prediction of][]{masunaga1998}. The accretion shock at the first core border remains supercritical throughout the simulations,
with the vast majority of the accretion energy being lost at the border in the form of radiation. The accretion shock at the second
core border was however found to be subcritical, with very little energy converted to radiation; the second core appears to absorb
all the energy from the infalling material. In addition, unlike the predictions of \citet{stahler1980} and the calculations of
\citet{schonke2011}, the dust destruction front located between the first and second core borders has only a very minimal effect on
the hydrodynamic properties of the pre-stellar system \citepalias[this was also reported in][]{tomida2013}. Additionally, we found
that once the first core is formed, less than $\sim 1000$ years go by before the second core is formed.

We also performed simulations of the collapse of a 0.1 and 10~\msun parent cloud, in order to confirm the robustness of the results 
stated above. The properties of the first and second cores (apart from the first core lifetimes) were found to be quasi-independent
the initial mass of the cloud for a same thermal to gravitational energy ratio; the second cores formed in our simulations all have
a radius of $3\times 10^{-3}$ AU, a mass of $\sim 10^{-3}~\msun$ and an entropy at the centre of
$\sim10^{9}~\text{erg~K}^{-1}~\text{g}^{-1}$. Finally, further simulations varying the size and temperature of the parent cloud
yielded virtually equivalent results, endorsing a fairly universal mass and size of the second core.

The grey approximation for radiative transfer appears to perform well in one-dimensional simulations of protostellar collapse. It
reproduces accurately the multi-frequency results, most probably because of the high optical thickness of the majority of the
protostar-envelope system. However, this multigroup method was developed primarily for 3D simulations where full spectral radiative
transfer is too heavy for current computational architectures. We still expect to see differences between grey and multigroup methods
in 3D due to different optical depths along different directions, parallel or perpendicular to the protostellar disk. In addition,
a simple estimate of the characteristic timescale of the second core suggests that the effects of using multigroup radiative transfer
may be more important in the long term evolution of the proto-star.

\acknowledgements

The research leading to these results has received funding from the European Research Council under the European Community's Seventh 
Framework Programme (FP7/2007-2013 Grant Agreement no. 247060). BC greatfully acknowledges support from the ANR Retour Postdoc
programme (ANR-11-PDOC-0031). Finally, the authors would like to thank the referee for very useful comments that have lead to a
more thorough analysis of our results which has in turn vastly improved the overall robustness of this work.

\bibliographystyle{aa}

\end{document}